\documentclass[journal]{IEEEtran}

\ifCLASSINFOpdf
\else
\fi

\usepackage{cite}
\usepackage{amsmath, amsfonts}

\usepackage[hidelinks]{hyperref}
\usepackage{url}

\usepackage{amsthm}
\newtheorem{theorem}{Theorem}

\usepackage{enumitem}

\usepackage{multirow}

\usepackage{booktabs}

\usepackage{subcaption}
\usepackage{xcolor}
\usepackage{color, colortbl}
\definecolor{GrayTop}{HTML}{e4e4e4}
\definecolor{GrayMid}{HTML}{FFA492}
\definecolor{AdCream}{HTML}{FCFCD4}
\definecolor{AdBlue}{HTML}{2e21c6}

\usepackage{graphicx}

\usepackage{nomencl}
\makenomenclature

\usepackage{etoolbox}
\renewcommand\nomgroup[1]{%
  \item[\bfseries
  \ifstrequal{#1}{I}{Indices, Sets and Vectors}{%
  \ifstrequal{#1}{V}{Variables}{%
  \ifstrequal{#1}{P}{Parameters}{}}}%
]}

\begin{document}
%
\title{Modelling Irrational Behaviour of Residential End Users using Non-Stationary Gaussian Processes}

\author{Nam Trong Dinh,~\IEEEmembership{Student Member,~IEEE,}
        Sahand Karimi-Arpanahi,~\IEEEmembership{Student Member,~IEEE,}
        Rui Yuan,~\IEEEmembership{Student Member,~IEEE,}
        S. Ali Pourmousavi,~\IEEEmembership{Senior Member,~IEEE,}
        Mingyu Guo, \\ Jon A. R. Liisberg, and Julian Lemos-Vinasco
}

\markboth{Accepted for publication in IEEE Transactions on Smart Grid}%
{Shell \MakeLowercase{\textit{et al.}}: Bare Demo of IEEEtran.cls for IEEE Journals}
%




\maketitle

\begin{abstract}
Demand response (DR) plays a critical role in ensuring efficient electricity consumption and optimal use of network assets. Yet, existing DR models often overlook a crucial element, the irrational behaviour of electricity end users. In this work, we propose a price-responsive model that incorporates key aspects of end-user irrationality, specifically loss aversion, time inconsistency, and bounded rationality. To this end, we first develop a framework that uses Multiple Seasonal-Trend decomposition using Loess (MSTL) and non-stationary Gaussian processes to model the randomness in the electricity consumption by residential consumers. The impact of this model is then evaluated through a community battery storage (CBS) business model. Additionally, we apply a chance-constrained optimisation model for CBS operation that deals with the unpredictability of the end-user irrationality. Our simulations using real-world data show that the proposed DR model provides a more realistic estimate of end-user price-responsive behaviour when considering irrationality. Compared to a deterministic model that cannot fully take into account the irrational behaviour of end users, the chance-constrained CBS operation model yields an additional 19\% revenue. Lastly, the business model reduces the electricity costs of solar end users by 11\%.
\end{abstract}


%
\IEEEpeerreviewmaketitle

\nomenclature[V,01]{\(x\)}{End-user consumption (kWh)}
\nomenclature[V,02]{\(x^+ / x^-\)}{End-user positive/negative net demand (kWh)}
\nomenclature[V,03]{\(x^\text{grid}\)}{End-user consumption from utility grid (kWh)}
\nomenclature[V,05]{\(G^\text{s}\)}{End-user spilled solar energy (kWh)}
\nomenclature[V,06]{\(\delta\)}{End-user solar credit offset (kWh)}
\nomenclature[V,07]{\(B\)}{End-user (dis)comfort function}
\nomenclature[V,08]{\(C\)}{End-user cumulative solar credits (kWh)}
\nomenclature[V,09]{\(\phi\)}{Binary variable for complementarity constraint}
\nomenclature[V,10]{\(P\)}{CBS dispatch power (kW)}
\nomenclature[V,11]{\(P^{\text{ch}} / P^{\text{dis}}\)}{CBS charging/discharging power (kW)}
\nomenclature[V,12]{\(E\)}{CBS state of energy (kWh)}
\nomenclature[V,13]{\(\upsilon^+ / \upsilon^-\)}{Community positive/negative net demand (kWh)}
\nomenclature[V,14]{\(\upsilon^\text{grid}\)}{Imported energy from the utility grid for charging CBS (kWh)}
\nomenclature[V,15]{\(\zeta^\text{peak}\)}{Community peak demand reduction (kW)}

\nomenclature[P,01]{\(\beta\)}{End-user price elasticity}
\nomenclature[P,02]{\(\hat{x}\)}{End-user originally expected consumption (kWh)}
\nomenclature[P,03]{\(\hat{x}^\text{hist}\)}{End-user historical consumption (kWh)}
\nomenclature[P,04]{\(\overline{x} / \underline{x}\)}{Upper/lower bounds of end-user energy consumption (kWh)}
\nomenclature[P,04]{\(\tilde{x}^+ / \tilde{x}^-\)}{End-user actual positive/negative net demand (kWh)}
\nomenclature[P,05]{\(\Delta x\)}{End-user adjusted energy consumption (kWh)}
\nomenclature[P,06]{\(\kappa\)}{End-user discounting degree}
\nomenclature[P,07]{\(\tau\)}{End-user discounting asymptote}
\nomenclature[P,07]{\(C^{\text{init}}\)}{End-user initial solar credits (kWh)}
\nomenclature[P,08]{\(\Delta t\)}{Length of time interval (h)}
\nomenclature[P,09]{\(\overline{\text{SoC}}/\underline{\text{SoC}}\)}{Upper/lower bounds of CBS state-of-charge (\%)}
\nomenclature[P,10]{\(E^{\text{cap}}\)}{CBS capacity (kWh)}
\nomenclature[P,11]{\(C^\text{rate}\)}{CBS C-rate}
\nomenclature[P,12]{\(E^{\text{cost}}\)}{CBS per-unit cost (\$/kWh)}
\nomenclature[P,13]{\(\Gamma\)}{CBS round-trip efficiency (\%)}
\nomenclature[P,14]{\(E^{\text{init}}\)}{CBS initial state-of-charge (kWh)}
\nomenclature[P,15]{\(\lambda^{\text{Opex}}\)}{CBS operation cost (\$/kW)}
\nomenclature[P,16]{\(T^e\)}{CBS expected lifetime (h)}
\nomenclature[P,17]{\(G\)}{Generated rooftop solar PV energy (kWh)}
\nomenclature[P,18]{\(\lambda^{\text{RT}}\)}{Real-time wholesale market price (\$/kWh)}
\nomenclature[P,19]{\(\lambda^{\text{Imp}}/ \lambda^{\text{Exp}}\)}{End-user import/export energy charge  (\$/kWh)}
\nomenclature[P,20]{\(\lambda^{\text{grid}}\)}{CBS operator grid usage charge (\$/kWh)}
\nomenclature[P,21]{\(\lambda^{\text{peak}}\)}{Peak demand incentive (\$/kWh)}
\nomenclature[P,22]{\(\lambda^{\text{fix}}\)}{DNSP fixed charge (\$/kW$\cdot$day)}
\nomenclature[P,23]{\(\lambda^{\text{cre}}\)}{Credit usage charge (\$/kWh)}
\nomenclature[P,24]{\(\lambda^{\text{rem}}\)}{Solar credits refund rate (\$/kWh)}
\nomenclature[P,25]{\(\eta_1, \eta_2, \eta_3\)}{Chance-constrained tolerance probability (\%)}
\nomenclature[P,26]{\(r\)}{Randomness component extracted from MSTL}
\nomenclature[P,27]{\(\tilde{r}\)}{Filtered randomness component}
\nomenclature[P,28]{\(f(\cdot)\)}{Gaussian processes random function}
\nomenclature[P,29]{\(m(\cdot)\)}{Gaussian processes mean function}
\nomenclature[P,30]{\(k_\text{SE}(\cdot,\cdot)\)}{Squared exponential kernel}
\nomenclature[P,31]{\(k(\cdot,\cdot)\)}{Non-stationary kernel}
\nomenclature[P,32]{\(\varepsilon\)}{Independent Gaussian noise}
\nomenclature[P,33]{\(\sigma_\text{SE}, l_\text{SE}\)}{Squared exponential kernel parameters}
\nomenclature[P,34]{\(\mu, \sigma^2\)}{Intervally filtered randomness mean/variance}
\nomenclature[P,35]{\(\mathbf{\Lambda}\)}{Square matrix of randomness components}
\nomenclature[P,36]{\(\mathbf{K_\text{SE}}\)}{Squared exponential covariance matrix}
\nomenclature[P,37]{\(\mathbf{K}\)}{Proposed covariance matrix}
\nomenclature[P,38]{\(x^{\text{rnd}}, x^{\text{rnd}\ast}\)}{Realised randomness consumption (kWh)}
\nomenclature[P,39]{\(\hat{x}^\text{rnd}\)}{Expected randomness consumption}
\nomenclature[P,40]{\(\hat{\mu}^\text{rnd}\)}{Expected randomness consumption mean}
\nomenclature[P,41]{\(\hat{\sigma}^{\text{rnd}^2}\)}{Expected randomness consumption variance}

\nomenclature[I,01]{\(n/N\)}{Index/Set of local prosumers}
\nomenclature[I,02]{\(\mathcal{H}\)}{Set of receding horizons}
\nomenclature[I,03]{\(\mathcal{T}\)}{Set of time intervals in a receding horizon}
\nomenclature[I,04]{\(\mathcal{T}^\text{RB}\)}{Set of time intervals in a rebound horizon}
\nomenclature[I,05]{\(d/\mathcal{D}\)}{Index/Set of operational day}
\nomenclature[I,06]{\(h,t,j,i\)}{Indices for time intervals}
\nomenclature[I,07]{\(R\)}{Set of randomness components}
\nomenclature[I,08]{\(\tilde{R}\)}{Set of filtered randomness components}
\nomenclature[I,09]{\(\mathbf{x}^{\text{rnd}}, \mathbf{x}^{\text{rnd}\ast}\)}{Vectors of realised randomness consumption}
\nomenclature[I,10]{\(\mathbf{\hat{x}}^{\text{rnd}}\)}{Vector of expected randomness consumption}
\nomenclature[I,11]{\(\mathcal{X}, \mathcal{X}^\ast, \mathcal{\hat{X}}\)}{Input vectors for non-stationary kernel}

\printnomenclature[0.655in]

\section{Introduction}
\subsection{Motivation}

\IEEEPARstart{D}{emand} response (DR) based on time-varying price signals provides many advantages to both electricity consumers and utility companies. 
Recognising this, various approaches have been proposed to model the consumption behaviour of residential users under different pricing schemes, namely time-of-use tariff, critical-peak pricing, and real-time (RT) wholesale market prices \cite{demand2012zhang, vardakas2015a, residential2013samuel, dinh2023cost}. DR frameworks generally employ a game-theoretic approach, assuming that electricity customers are rational players that act in the interest of minimising electricity bills by adjusting their consumption to price signals \cite{a2016yu, pricing2021werner, dinh2023cost}. However, this rational behaviour, while typically observed in the short term, may not always hold over extended periods. Behavioural economics provides a framework for understanding this deviation from rational decision-making, recognising that residential users often lack complete rationality in the actual decision-making processes \cite{behavioral2016selahattin}. As a result, neglecting such irrational behaviour can lead to ill-informed decisions with significant financial consequences \cite{elisha2015household}. 

To understand the impact of end-user non-ideal behaviour on economic benefits, this paper examines the business model of community battery storage (CBS). The rapid adoption of rooftop solar photovoltaic (PV) systems in Australia has resulted in the emergence of local energy communities that are increasingly considering the incorporation of CBS as an essential element \cite{AusgridCBS}. Recently, the Australian federal government has announced the allocation of \$224.3 million to community battery schemes \cite{cbsscheme}, which could see the installation of more than 400 community batteries throughout the country in the next few years \cite{cbsscheme}. In preparation for the rollout of new medium-scale battery systems in distribution networks, many Australian distribution network service providers (DNSPs) have introduced trial network tariffs to promote the use of CBS to support their networks \cite{AERtrial}.

This presents an opportunity for a new business model with respect to CBS. Therefore, it is imperative to perform a business analysis to evaluate the viability of these investments. A crucial factor that influences the profitability of CBS projects is the ability to forecast daily electricity consumption within the local energy community. This is important for two reasons: first, to optimally engage in energy arbitrage, which serves as one of the revenue streams of CBS \cite{optimal2022Dinh}; second, to leverage CBS for managing peak power demand in distribution networks, which is the main driver of costly network upgrades for DNSPs \cite{residential2013samuel, peakdemandissue}. 
Thus, the success of a CBS project depends greatly on the ability to forecast local electricity consumption, which, in turn, is directly related to understanding consumers' behaviour.

\subsection{Related works}

Various machine learning (ML) forecasting models have been developed to predict daily consumption, which can be deterministic or probabilistic forecasts \cite{probabilistic2021julian, lemosvinasco2022economic}. However, they all share one fundamental challenge; trained models with historical data may systematically fail to produce accurate forward-looking forecasts if future consumption patterns deviate significantly from historical data. This is particularly crucial for modern electricity end users, 
who are increasingly equipped with rooftop PV, electric vehicles and home energy management systems (HEMS). Due to their reliance on various uncertain factors, these advances can introduce substantial variability into the net demand profile. This change is exacerbated by the highly dynamic pricing environment of RT wholesale prices, which can be offered directly to residential customers, as has been done by a few retailers, e.g., Amber Electric in Australia \cite{amberelectric}. For this reason, various price-responsive models have been proposed to estimate the consumption behaviour of residential users for analysing new business models. In most studies, the time-varying DR programs are encapsulated within a nonlinear utility function \cite{optimal2022Dinh, a2016yu, peer2022jiang, dinh2023cost}. This function considers the linear cost of electricity along with a nonlinear (dis)comfort function that depends on electricity usage. The incorporation of the (dis)comfort function facilitates the modelling of nonlinear decision-making, reflecting the complex ways in which electricity customers react to price signals. Furthermore, the function mimics the principle of loss aversion, that is, a psychological property in behavioural economics that describes irrational behaviour of the end users \cite{thirty2013barberis}. In loss aversion theory, human perception is modelled asymmetrically, with individuals being influenced more by the fear of losses than by the potential gains. Unlike loss aversion \cite{accomodating2019radoszynski}, other properties of behavioural economics, which will be detailed in the following paragraph, are not mathematically modelled in the DR literature.

In \cite{behavioral2016selahattin}, different properties of irrational behaviour were recommended for DR programs, including but not limited to bounded rationality and time inconsistency. Although bounded rationality has been considered in various studies, its definition and interpretation lack consistency. For example, in \cite{accomodating2019radoszynski}, the authors addressed the bounded rationality by assuming that end users are unaware of the financial advantages of DR, thus removing the financial aspect from the formulation of the problem. In another study \cite{electric2021marin}, while financial incentives are retained, the authors deliberately perturb optimal solutions using normally distributed noise, without any clear definition of ``noise''. In contrast to bounded rationality, time inconsistency is rarely considered in the literature on DR. The theory proposes that people's inter-temporal choices are inconsistent over time largely due to their inherent bias towards immediate outcomes compared to those in the future \cite{models2007wilson}. This bias can be modelled using a hyperbolic discounting function rather than an exponential discount rate, commonly seen in economic theory \cite{steel2007nature}. A well-known application of time inconsistency is the explanation of the procrastination phenomenon \cite{behavior2010hunt}. In \cite{measuring2014shuling}, a survey of residential electricity customers revealed procrastination as a barrier to energy-saving behaviour. The study pointed out that the intangible discomfort associated with energy-saving activities often leads people to postpone planned DR tasks. Despite the importance of this issue, there has not yet been a mathematical formulation of time inconsistency in the context of DR.

\subsection{Objectives and contributions}

In this paper, we integrate various aspects of irrational behaviour, namely loss aversion, time inconsistency, and bounded rationality, into the formulation of the DR model. We define a DR activity as a load shift. 
In our approach, loss aversion is considered in the (dis)comfort model for load shifting such that end users experience significant discomfort with a reduction in demand and a comparatively smaller increase in comfort when increasing consumption. Time inconsistency is addressed by discounting the (dis)comfort values for intervals that are far ahead of the current time step. Hence, we leverage receding horizon operation (RHO) to illustrate the changing decisions of end users over time, demonstrating the effect of procrastination on their decision-making process. Bounded rationality is considered as the cognitive limits of actual consumption, causing end users to opt for suboptimal solutions. These cognitive constraints originate from the inherent randomness of end users' behaviour, of which they are unaware during decision making. To this end, we propose a framework to determine the time-dependent randomness of consumption using Multiple Seasonal-Trend decomposition using Loess (MSTL) \cite{MSTL2021kasun}. The framework then leverages Gaussian processes with non-stationary kernel (see \cite{advanced2022marcus}) to generate randomness in users' DR activities. Having awareness of the irrationality of end users, the CBS operator applies a chance-constrained optimisation for battery operation \cite{energy2017baker, su2022bi, marasciuolo2023chance}, where local consumption is considered as the stochastic parameter to effectively manage uncertainty. In summary, this paper offers three main contributions:
\begin{itemize}[wide]
    \item Mathematical modelling of electricity end users' irrational behaviour in the DR model that integrates time inconsistency and bounded rationality with established loss aversion models.
    \item Proposing a framework that models end users' randomness in electricity consumption by leveraging MSTL and Gaussian processes.
    \item Developing an adaptive non-stationary kernel for Gaussian processes, capturing the varying magnitudes of consumption randomness across different times of the day.
\end{itemize}

The remainder of this paper is organised as follows. In section \ref{sec:lec}, we outline the structure of the local energy community with CBS. Then, we delve into the mathematical modelling of local end users and CBS operator in sections \ref{sec: end-users_optimisation} and \ref{sec:CBS_operator}, respectively. The simulation results are discussed in section \ref{sec:sim_study}. Lastly, we conclude the paper in section \ref{sec:conc}.

\section{Local energy community with CBS}
\label{sec:lec}

\begin{figure}[!t]
    \centering
    \includegraphics[trim={1.45cm 7.1cm 1.45cm 7.3cm}, clip, width=\linewidth]{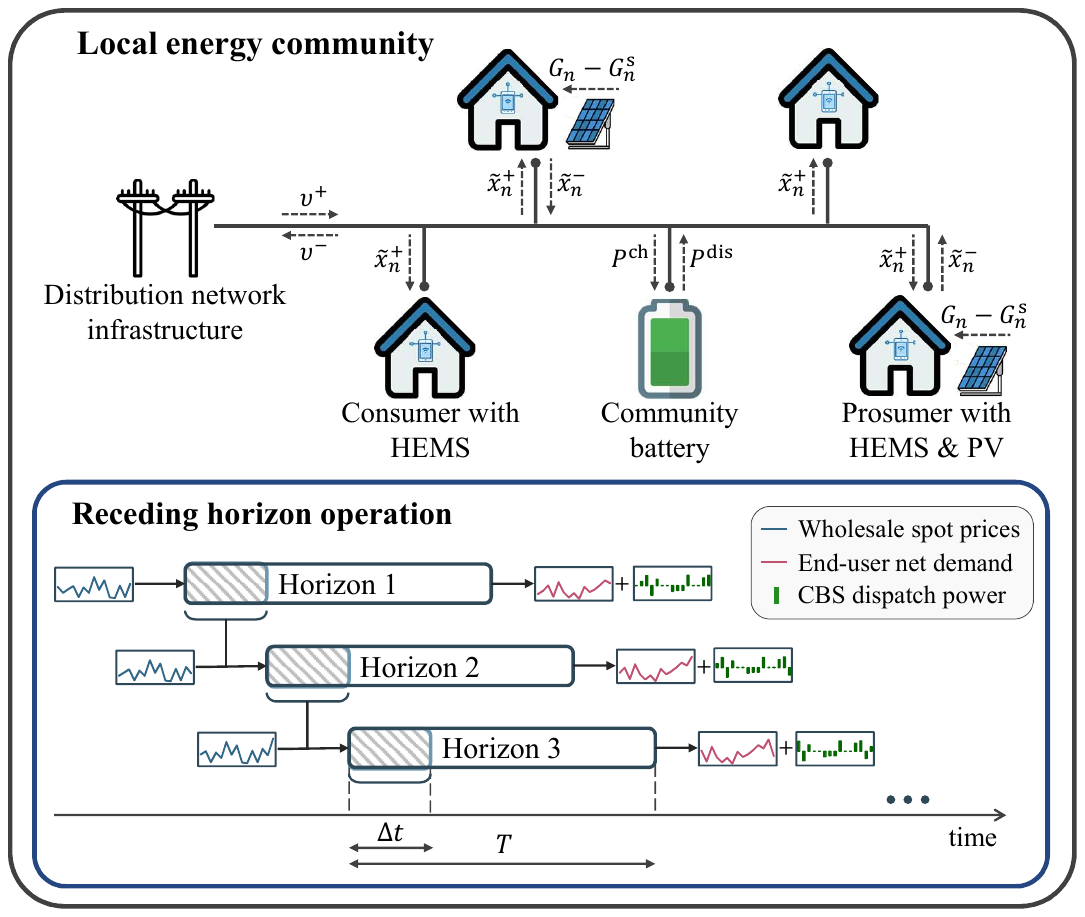}
    \caption{Local energy community structure with RHO setting}
    \label{fig:cbs_flowchart}
\end{figure}

The upper half of Fig. \ref{fig:cbs_flowchart} presents the structure of a local energy community. In this setup, the CBS operator functions as an electricity retailer that passes on RT wholesale market prices to end users in the local energy community. End users can be classified into two groups: those with rooftop solar PV, referred to as solar prosumers, and those without it, referred to as consumers. More than 33\% of Australian households have adopted rooftop solar PV \cite{APVIsolarpv}. This substantial uptake has resulted in distribution networks often experiencing reverse power flow during midday. To manage this, many DNSPs have introduced flexible export limits, which could reduce the amount of solar energy that can be exported to the grid to as low as 1.5 kW \cite{SAPNflexibleexport}. However, this strategy leads to significant energy spillage from rooftop PV systems.
To make use of the otherwise wasted solar generation, the CBS operator can offer local prosumers to store their exported energy in the CBS in exchange for solar credits. These solar credits can later be used by prosumers to offset their electricity consumption at night. Furthermore, to increase profits, the CBS operator can sell the abundant electricity in the battery to local consumers at RT wholesale prices. Hence, the CBS operator can perform energy arbitrage by purchasing electricity from the utility grid at low prices and selling in the local community when prices are high. 
To demonstrate the irrational behaviour of the end users, we leverage RHO, as shown in the lower half of Fig.~\ref{fig:cbs_flowchart}, and consider a perfect forecast of RT prices. If end users have complete knowledge of future RT prices and are fully rational, which is often claimed in the existing DR models, they can optimise their DR strategies and adhere to planned consumption, regardless of the planning horizon. However, in reality, the irrationality of human behaviour can lead to changing decisions over time and, as a result, deviating from optimal consumption. 

The operation of the local energy community can be summarised as follows. Using HEMS, end users optimise their day-ahead consumption in accordance with their load-shifting behaviour to take advantage of RT price fluctuations. The planned consumption, as predicted by HEMS, is then reported to the CBS operator for optimal scheduling of CBS (dis)charging. Under the RHO regime, the process is repeated consecutively in subsequent receding horizons, as shown in Fig. \ref{fig:cbs_flowchart}. Note that due to the change in behaviour and randomness of end users from time inconsistency and bounded rationality properties, consumption information for the same time interval can vary significantly across different receding horizons. To deal with unexpected changes in end users behaviour at the time of consumption, the CBS operator solves a stochastic optimisation to manage the CBS. This model contrasts with a deterministic model, which neglects the uncertainties associated with consumption patterns.

\section{Problem formulation for local end users}
\label{sec: end-users_optimisation}

In this section, we first develop the utility model for local end users that integrates the loss aversion and time inconsistency properties of irrational behaviour. Following this, we elaborate on bounded rationality by constructing a framework to depict end users' randomness, leveraging MSTL and non-stationary Gaussian processes.

\subsection{Utility model of local end users with loss aversion}
\label{subsec:loss_aversion}

We denote $N = \{1,2,\dots,|N|\}$ as the set of local end users and $\mathcal{H} = \{1,2,\cdots,H\}$ as the set of receding horizons. For each receding horizon $h \in \mathcal{H}$, the optimisation model is solved for a one-day lookahead horizon, where the corresponding set of time intervals is given by $\mathcal{T} = \{1,2,\cdots,T\}$. The optimisation model of the end users for each receding horizon $ h $ is formulated as follows:
\begin{subequations}
\label{eqn:pro_sol}
\begin{flalign}
\label{eqn:min_uti_pro}
    \min_{\mathbf{\Psi_{n,h}}} U_{n,h} = \sum_{t \in \mathcal{T}} & \big[ \lambda^{\text{RT}}_{h,t} x^{\text{grid}}_{n,h,t} + \lambda^{\text{Imp}}_{h,t} x^+_{n,h,t} + \lambda^{\text{Exp}}_{h,t} x^-_{n,h,t} \nonumber \\
    & + B(x_{n,h,t}) \big] \quad \forall n \in N, \; \forall t \in \mathcal{T} &&
\end{flalign}
where
\begin{flalign}
\label{eqn:pro_discomfort}
    B(x_{n\!,h\!,t})\! = \! -\!\lambda^{\text{Ref}}_h \! \left[\!1 \!+\! \frac{(x_{n\!,h\!,t} \!-\! \hat{x}_{n\!,h\!,t})}{2\beta_{n,h,t}\hat{x}_{n,h,t}}\!\right]\!\! (x_{n,h,t}\! -\! \hat{x}_{n,h,t}) &&
\end{flalign}
s.t.
\begin{flalign}
\label{eqn:rebound_const}
    \sum_{t \in \mathcal{T}^\text{RB}} x_{n,h,t} = \!\!\! \sum_{t \in \mathcal{T}^\text{RB}} \hat{x}_{n,h,t} + \Delta x_{n,h} \quad \forall n \in N, &&
\end{flalign}
\begin{flalign}
\label{eqn:cons_boundary}
    \underline{x}_{n,h,t} \leq x_{n,h,t} \leq \overline{x}_{n,h,t} \quad \forall n \in N, \; \forall t \in \mathcal{T}, &&
\end{flalign}
\begin{flalign}
\label{eqn:prosumer_net}
    x_{n,h,t} \! - \! G_{n,h,t} + G^\text{s}_{n,h,t} = x^+_{n,h,t} \! - x^-_{n,h,t} \quad \forall n \! \in \! N, \; \forall t \! \in \! \mathcal{T}, &&
\end{flalign}
\begin{flalign}
\label{eqn:pos_net_restriction}
    x^+_{n,h,t} \leq M \cdot (1 - \phi_{n,h,t}) \quad \forall n \in N, \; \forall t \in \mathcal{T}, &&
\end{flalign}
\begin{flalign}
\label{eqn:neg_net_restriction}
    x^-_{n,h,t} \leq M \cdot \phi_{n,h,t} \quad \forall n \in N, \; \forall t \in \mathcal{T}, &&
\end{flalign}
\begin{flalign}
\label{eqn:spilled_solar}
    G^\text{s}_{n,h,t} \leq G_{n,h,t} \quad \forall n \in N, \; \forall t \in \mathcal{T}, &&
\end{flalign}
\begin{flalign}
\label{eqn:credit_offset}
    x^+_{n,h,t} - \delta_{n,h,t} = x^{\text{grid}}_{n,h,t} \quad \forall n \in N, \; \forall t \in \mathcal{T}, &&
\end{flalign}
\begin{flalign}
\label{eqn:cumulative_credit}
    C_{n,h,t} \!=\! C^{\text{init}}_{n,h} + \!\! \sum_{j=1}^t (x^-_{n,h,j} \!-\! \delta_{n,h,j}) \quad \forall n \! \in N, \, \forall t \!  \in \mathcal{T}, &&
\end{flalign}
\end{subequations}
where $\mathbf{\Psi_{n,h}} = \{x_{n,h,t}, x^+_{n,h,t}, x^-_{n,h,t}, \phi_{n,h,t}, \delta_{n,h,t}, G^\text{s}_{n,h,t},$ $x^{\text{grid}}_{n,h,t}, C_{n,h,t}\}$. The objective function in \eqref{eqn:min_uti_pro} aims to minimise the electricity cost and the discomfort incurred from load shifting activities. Unlike conventional retailers that bundle electricity prices with network charges \cite{bundletariff}, end users who opt for RT wholesale market prices often see a more transparent cost breakdown \cite{amberelectric}. In our model, cost of electricity is broken down into three components, namely energy usage charges, network import charges, and network export charges as specified by the first three terms in \eqref{eqn:min_uti_pro}. Although solar PV export is generally encouraged, reverse power flow can lead to over-voltages and congestion in distribution network feeders with high uptake of PV. For this reason, the network export charge, $\lambda^{\text{Exp}}_{h,t}$, in our model can be configured to positive or negative values, depending on the DNSP, to discourage or encourage solar export, respectively \cite{AusrgidCBStariff}. The quadratic discomfort function in \eqref{eqn:pro_discomfort} is a descending convex function that captures the loss aversion behaviour of end users. When they decrease consumption from their expected value, i.e., $x_{n,h,t} < \hat{x}_{n,h,t}$, the function generates a positive discomfort value. Conversely, when end users increase their consumption, they experience negative discomfort, implying a state of satisfaction. However, a specific amount of negative deviation from $\hat{x}_{n,h,t}$ would result in greater discomfort compared to the comfort value gained for the same amount of positive deviation from $\hat{x}_{n,h,t}$. This illustrates the asymmetry in psychological behaviour with respect to loss and gain, which is central to loss aversion. In addition to expected consumption, the discomfort function in \eqref{eqn:pro_discomfort} incorporates external factors, represented by a price reference, $\lambda^{\text{Ref}}_h := \max\{\lambda^{\text{RT}}_{h,t} + \lambda^{\text{Imp}}_{h,t}| t \in \mathcal{T}\}$, and a time-varying price elasticity. In \cite{price2020zsuzanna}, empirical evidence suggests that the price elasticity of residential electricity consumption is relatively low, with $|\beta_{n,h,t}| < 1$.

Since DR activity is performed via load shifting, there is a rebound effect that must be considered. This implies that any reduction (increase) in demand must be accompanied by an increase (reduction) in consumption within a short period of time, as specified in \eqref{eqn:rebound_const}. We denote the rebound horizon as $\mathcal{T}^\text{RB} \!=\! \{1,\!2,\!\cdots\!,\!T^\text{RB}\} \! \subset \! \mathcal{T}$. Due to the sequential solving of the RHO, only the decision variables in the first interval of each receding horizon are binding. Therefore, to satisfy the rebound effect, consumption deviation, $\Delta x_{n,h} \!=\! \sum_{j=1}^{h-1} \big( \hat{x}_{n,j} - x_{n,j} \big)$, from previous receding horizons is incorporated into the current receding horizon. Constraint \eqref{eqn:cons_boundary} defines the boundaries of consumption in each interval. Constraint \eqref{eqn:prosumer_net}--\eqref{eqn:neg_net_restriction} represents the net demand of end users with respect to consumption, solar generation, and solar energy spill. In particular, constraint \eqref{eqn:prosumer_net} separates net demand into positive and negative net demand. For local consumers without a rooftop PV system, solar generation is zero. Hence, only $x^+_{n,h,t}$ can have non-zero values. However, for solar prosumers, depending on the amount of solar generation, the net demand can be positive or negative. Normally, the optimisation solver would prefer to set $x^+_{n,h,t}$ or $x^-_{n,h,t}$ to zero due to the charges on the import and export of energy in \eqref{eqn:min_uti_pro}. However, the high penetration of renewable energy sources (RES) has frequently driven RT prices below zero \cite{negativepricing}, which can make optimisation unbounded (due to $x^+_{n,h,t}$ and $x^-_{n,h,t}$ increasing indefinitely). To avoid this issue, we consider complementarity constraints in \eqref{eqn:pos_net_restriction} and \eqref{eqn:neg_net_restriction}. Moreover, when the RT prices are negative and the network export charge is positive, the favourable reaction of the prosumers is to curtail solar generation. To model this behaviour, we consider the spilled solar energy in \eqref{eqn:spilled_solar}, which can be managed at the inverter level with the help of HEMS. Constraint \eqref{eqn:credit_offset} determines the electricity consumption of the utility grid by end users, after considering the offset of solar credits. Constraint \eqref{eqn:cumulative_credit} calculates the accumulated solar credits, where $C^{\text{init}}_{n,h} \!=\! C_{n,h-1,t=1}$. Note that while \eqref{eqn:spilled_solar}--\eqref{eqn:cumulative_credit} only affect solar prosumers, these constraints can also be applied to consumers without loss of generality. 

\subsection{Time inconsistency}
\label{subsec:time_inconsis}

Although the current discomfort function successfully captures the asymmetry in loss and gain, it cannot model the inconsistency of human actions in relation to intertemporal choices. For example, consider a receding horizon starting at 08:00, where end users can estimate that it is in their best financial interest to perform DR at 20:00 by reducing their load. Given the significant temporal distance (i.e., 12 hours), the end users might underestimate the discomfort incurred from the activity. However, as the time approaches 20:00, the end users become more aware of their comfort, leading them to choose to maintain the consumption level and avoid reducing their load. To account for this changing behaviour, we leverage the hyperbolic discounting function to model the distorted discomfort values over time \cite{steel2007nature}. The objective function in \eqref{eqn:min_uti_pro} is thus modified as follows:
\begin{flalign}
\label{eqn:time_incons}
    \min_{\mathbf{\Psi_{n,h}}} U_{n,j} = & \sum_{t \in \mathcal{T}} \big[ \lambda^{\text{RT}}_{h,t} x^{\text{g}}_{n,h,t} + \lambda^{\text{Imp}}_{h,t} x^+_{n,h,t} + \lambda^{\text{Exp}}_{h,t} x^-_{n,h,t} \nonumber \\
    & + \! \frac{1 \!+\! \tau \! \cdot \! t\! \cdot \! \kappa_n}{1 \! + \! t \! \cdot \! \kappa_n} B(x_{n,h,t}) \big] \quad \forall n \! \in \! N, \, \forall t \! \in \! \mathcal{T}, &&
\end{flalign}
where $\kappa_n$ depicts the discounting degree and varies for different end users, and $\tau \! \in \! [0, \!1]$ is the horizontal asymptote of the hyperbolic discounting function. The asymptote is necessary to avoid the function approaching zero for large $t$. As $t$ extends into the future, representing a longer lookahead horizon, lower values of $\tau$ and higher values of $\kappa_n$ significantly enhance the hyperbolic discounting effect. This results in differences (errors) between estimated and actual consumption, reflecting a similar increase in uncertainty as observed in ML forecasting models. In such models, forecasts accuracy typically decreases as the prediction horizon extends \cite{probabilistic2021julian, CORNELL2024a, kumar2022optimal}. Although humans apply discounting to financial benefits \cite{steel2007nature}, we consider that the financial discount is negligible for one day ahead.

\subsection{Consumption randomness as bounded rationality}
\label{subsec:bounded_rationality}

\begin{figure}[!t]
    \centering
    \includegraphics[trim={5.0cm 9.5cm 5.0cm 9.5cm}, clip, width=0.84\linewidth]{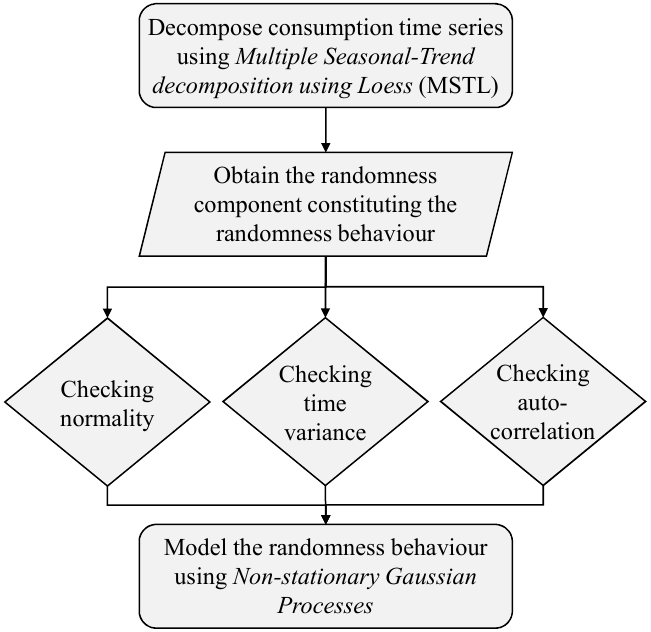}
    \caption{A flowchart summarising end-user consumption randomness modelling}
    \label{fig:randomness_process}
\end{figure}

By solving the end users' optimisation model in \eqref{eqn:time_incons} and \eqref{eqn:pro_discomfort}--\eqref{eqn:cumulative_credit}, the HEMS can operate household appliances automatically and provide the price-responsive consumption information to the CBS operator. However, in reality, end users can still deviate significantly from the estimation by HEMS, e.g., by manually changing the appliance settings \cite{residential2013samuel}. We interpret this deviation as random behaviour, a component of the consumption time series that is difficult to predict. Consider the variables with ($\ast$) as the optimised values committed from previous receding horizons, the actual net demand is thus represented as:
\begin{align}
\label{eqn:actual_consumption}
    \tilde{x}^+_{n,h} - \tilde{x}^-_{n,h} = x^\ast_{n,h} + x^\text{rnd}_{n,h} - G_{n,h} + G^{\text{s}\ast}_{n,h}, &&
\end{align}
where $x^\text{rnd}_{n,h}$ is the consumption randomness resulting from the bounded rationality of the end users. In this model, we consider $x^\text{rnd}_{n,h}$ as the sample from a Gaussian process given input $\mathcal{X}_{n,h}$, and independent Gaussian noise as follows:
\begin{align}
\label{eqn:randomness_realisation}
    x^\text{rnd}_{n,h} = f_n(\mathcal{X}_{n,h}) + \varepsilon_{n,h}, &&
\end{align}
where $f_n \! \sim \! \mathcal{GP}(m_n (\cdot), k_n(\cdot, \cdot))$ and $\varepsilon_{n,h} \! \sim \! \mathcal{N}(0, \sigma^2_{\varepsilon_{n,h}})$. Here, $k_n(\cdot, \cdot)$ is the proposed non-stationary kernel, which is configured using the randomness behaviour of the end users. Figure \ref{fig:randomness_process} shows a graphical summary of the end-user consumption randomness modelling.

\subsubsection{MSTL decomposition on consumption data}
To extract the random behaviour, we apply the MSTL decomposition technique to historical consumption. The technique extends the traditional Seasonal-Trend decomposition using Loess (STL) to allow for capturing multiple seasonal patterns in a time series. By considering multiple seasonalities during decomposition, our aim is to ensure that the extracted randomness component encapsulates the lowest degree of predictability \cite{kani2020improving}. The MSTL decomposition is expressed as follows:
\begin{align}
\label{eqn:MSTL}
    \hat{x}^\text{hist}_{n,h} = S^\text{daily}_{n,h} + S^\text{weekly}_{n,h} + T_{n,h} + r_{n,h}, &&
\end{align}
where $S^\text{daily}_{n,h}$ and $S^\text{weekly}_{n,h}$ denote the daily and weekly seasonal components; $T_{n,h}$ is the trend component; and $r_{n,h}$ is the randomness component constituting the randomness behaviour of end users.

\subsubsection{Normality check of the randomness component}
Since the consumption randomness in \eqref{eqn:randomness_realisation} is generated using the Gaussian process, it is necessary to ensure that the extracted randomness behaviour in \eqref{eqn:MSTL} follows a Gaussian distribution. In particular, we perform the normality check for each interval of the day using the Kolmogorov–Smirnov (KS) test \cite{the1951frank}. Let $d \in \mathcal{D} = \{1,2,\cdots,D\}$ represent each day in the simulation dataset, with $D = \frac{H}{T}$ being the total number of days. Since $t \in \mathcal{T}$ can be considered as the interval within a day, we can rewrite the randomness component $r_{n,h}$ as:
\begin{align}
\label{eqn:randomness_comp}
    r_{n,h} = r_{n,t,d} \quad \text{where} \quad h = (d-1)T + t. &&
\end{align}

To create a distribution for each interval within a day, we denote $R_{n,t} = \{r_{n,t,d} | d \in \mathcal{D}\}$ as the set of all randomness values corresponding to the interval $t$ across all days $d \in \mathcal{D}$. The KS test is then applied to each set $R_t$, testing the null hypothesis $\mathbf{H}_0$: $R_t$ is normally distributed.
However, due to many outliers, the KS test often rejects the null hypothesis. For this reason, we use the interquartile range (IQR) thresholding strategy to remove outliers \cite{mishra2019descriptive}. Although undesired for the analysis, the outliers are part of the randomness behaviour. Therefore, we consider the removed outliers to be encapsulated under the independent Gaussian noise $\varepsilon_{n, h}$ in \eqref{eqn:randomness_realisation}. Overall, the KS test is applied to the filtered randomness component represented as $\tilde{R}_{n,t} \! = \! \{\tilde{r}_{n,t,d} | d \! \in \! \mathcal{D}\}$. We then fit a normal distribution on each set $\tilde{R}_{n,t}$ to obtain the mean $\mu_{n,t}$ and variance $\sigma^2_{n,t}$ such that $\! \tilde{r}_{n,t,d} \! \sim \! \mathcal{N}(\mu_{n,t}, \sigma_{n,t}^2)$. Given that the randomness component is trend adjusted in \eqref{eqn:MSTL}, we have $\mu_{n,t} \! \approx \! 0$. Consequently, we consider a zero-mean prior $m_n(\cdot) \! = \! 0$ for $f_n$. On the other hand, variance $\sigma^2_{n,t}$ becomes important as a measure of randomness in the consumption of end users.

\subsubsection{Autocorrelation of the randomness component}
Although MSTL effectively removes multiple seasonal patterns, the remaining randomness component may still exhibit autocorrelated behaviour. This happens because of the faster evolving seasonality that was overlooked when constructing the model, e.g., the hourly seasonality. Imagine a situation where the randomness of consumption arises from the unexpected usage of household appliances by end users. The effect of such actions can last for a few hours until the appliances complete their cycle. For this reason, autocorrelation within the randomness component would be expected. In this study, we use the autocorrelation function (ACF) plot to visually show the autocorrelation of the randomness component.

\subsubsection{Non-stationary Gaussian processes}
The established autocorrelation and the normal distribution of the randomness behaviour allow for the modelling of Gaussian processes. Fundamentally, the design of Gaussian processes is based on the design of the kernel (or covariance function) $k_n$ \cite{görtler2019a}. The most common kernel is the squared exponential (SE) kernel \cite{advanced2022marcus}. For a univariate $\! h$, the SE kernel $\! k_\text{SE} \!: \! \mathbb{R} \! \times \! \mathbb{R} \! \mapsto \! \mathbb{R}$ is denoted as:
\begin{align}
\label{eqn:SE_kernel}
    k_\text{SE}(h_i, h_j) = \sigma_\text{SE}^2 \exp{\bigg(-\frac{(h_i - h_j)^2}{2l_\text{SE}^2} \bigg)}. &&
\end{align}

The SE kernel is stationary  inherently because it depends only on the temporal distance between two time intervals, not the intervals themselves. This poses a challenge in modelling the randomness of end users, as it implies constant randomness behaviour throughout the day, which is unrealistic. Hence, we extend the stationary SE kernel to a non-stationary kernel by incorporating the aforementioned randomness measurement $\sigma^2_{n,t}$. To effectively incorporate this measure into the kernel function, we first set $\sigma^2_\text{SE}$ at a constant value of $1$. This adjustment allows $\sigma^2_{n,t}$ to function as the sole variance parameter, ensuring that our adapted non-stationary kernel accurately represents the varying randomness of end-user consumption. We define $\! \mathbf{K_\text{SE}} = [k_\text{SE}(h_i, h_j)]_{i,j=1}^{H}$ as the $\! H \times H \!$ covariance matrix constructed from the SE kernel, and $\mathbf{K}_n \in \mathbb{R}^{H \times H}$ the covariance matrix from the non-stationary kernel $k_n$. It is imperative to ensure that $\mathbf{K}_n$ is positive semi-definite.

\begin{theorem}
Let $\mathbf{\Lambda}$ be a square diagonal matrix and $\mathbf{K}$ be a positive semi-definite matrix, then $\mathbf{\Lambda}^\top \mathbf{K} \mathbf{\Lambda}$ is symmetric and positive semi-definite.
\end{theorem}
In this model, we consider $\mathbf{K}_n = \mathbf{\Lambda}_n^\top \mathbf{K_\text{SE}} \mathbf{\Lambda}_n$ where $\mathbf{\Lambda}_n \in \mathbb{R}^{H \times H}$ is represented as:
\begin{align}
\label{eqn:randomness_matrix}
\mathbf{\Lambda}_n \! = \! \begin{bmatrix}
\text{diag}(\sigma_{n,1}, \! \cdots \!, \sigma_{n,T})_1 & & \\
& \!\!\!\!\!\! \ddots \!\!\!\!\!\! & \\
& & \text{diag}(\sigma_{n,1}, \! \cdots \!, \sigma_{n,T})_D
\end{bmatrix}
\end{align}
Overall, given the established zero-mean function and the non-stationary kernel $k_n \!: \mathbb{R}^2 \times \mathbb{R}^2 \mapsto \mathbb{R}$, we have $\mathbf{x}^\text{rnd}_{n}|\mathcal{X}_n \sim \mathcal{N}(0, \mathbf{K}_n + \sigma^2_{\varepsilon_n}I)$ follows a multivariate Gaussian distribution, where $\mathbf{x}^\text{rnd}_{n} \! = \! (x^\text{rnd}_{n,1}, \! \cdots \!, x^\text{rnd}_{n,H})^\top$ and $\mathcal{X}_n = [ (h, \sigma_{n,h}) \! : h \! \in \! \mathcal{H}]^\top$.

\section{Problem formulation for the CBS operator}
\label{sec:CBS_operator}

\subsection{Considering loss aversion and time inconsistency via a deterministic model}

Upon receiving the optimised consumption from the end users for each receding horizon, the CBS operator runs its optimisation problem to schedule the CBS operation, which is given below for each receding horizon $h \in \mathcal{H}$:
\begin{subequations}
\label{eqn:CBS_operation}
\begin{flalign}
\label{eqn:CBS_profit}
    \min_{\mathbf{\Psi^{\text{CBS}}_h}} O^{\text{CBS}}_h = & \sum_{t \in \mathcal{T}} \! \left( \lambda^\text{RT}_{h,t} \upsilon^+_{h,t} + \lambda^{\text{grid}} \upsilon^{\text{grid}}_{h,t}  + \! \lambda^{\text{Opex}} P^\text{ch}_{h,t} \Delta t \right) \nonumber \\
    & - \lambda^\text{peak}_h (\zeta^\text{local}_h - \zeta^\text{user}_h) &&
\end{flalign}
s.t.
\begin{flalign}
\label{eqn:net_demand}
    \upsilon^+_{h,t} \! - \! \upsilon^-_{h,t}  = \sum_{n \in N} \!\! \big( x^+_{n,h,t} - x^-_{n,h,t} \big) + P_{h,t} \Delta t \quad \forall t \! \in \! \mathcal{T}, &&
\end{flalign}
\begin{flalign}
\label{eqn:peak_definition}
    \zeta^\text{local}_h \geq \upsilon^+_{h,t} \quad \forall t \in \mathcal{T}, &&
\end{flalign}
\begin{flalign}
\label{eqn:complementarity_local_net}
    0 \leq \upsilon^+_{h,t} \perp \upsilon^-_{h,t} \geq 0 \quad \forall t \in \mathcal{T}, &&
\end{flalign}
\begin{flalign}
\label{eqn:battery_soc}
    E_{h,t} = E^{\text{init}}_h + \sum_{j=1}^t \big( P^{\text{ch}}_{h,j} - \frac{1}{\Gamma} P^{\text{dis}}_{h,j}\big) \Delta t \quad \forall t \in \mathcal{T}, &&
\end{flalign}
\begin{flalign}
\label{eqn:charging_power}
    P_{h,t} = P^{\text{ch}}_{h,t} - P^{\text{dis}}_{h,t} \quad \forall t \in \mathcal{T}, &&
\end{flalign}
\begin{flalign}
\label{eqn:charging_limit}
    - E^{\text{cap}} C^\text{rate} \leq P_{h,t} \leq E^{\text{cap}} C^\text{rate} \quad \forall t \in \mathcal{T}, &&
\end{flalign}
\begin{flalign}
\label{eqn:soc_boundary}
    \underline{\text{SoC}} \, E^{\text{cap}} \leq E_{h,t} \leq \overline{\text{SoC}} \, E^{\text{cap}} \quad \forall t \in \mathcal{T}, &&
\end{flalign}
\begin{flalign}
\label{eqn:LUoS}
    \upsilon^{\text{grid}}_{h,t} \geq P^{\text{ch}}_{h,t} \Delta t - \sum_{n \in N} x^-_{n,h,t}  \quad \forall t \in \mathcal{T}, &&
\end{flalign}
\end{subequations}
where $\mathbf{\Psi^{\text{CBS}}_h} \!=\!\! \{ E_{h,t}, P^{\text{ch}}_{h,t}, P^{\text{dis}}_{h,t}, \upsilon^+_{h,t}, \upsilon^-_{h,t}, \upsilon^{\text{grid}}_{h,t}, \zeta^\text{local}_h\}$. The objective of the CBS operator in \eqref{eqn:CBS_profit} is to minimise the net operating cost. This cost includes three components: 1) energy usage charges from the wholesale market, 2) network usage charges when charging from the utility grid, and 3) throughput cost considered as the operating expense (Opex) of the CBS. Regarding network usage charges, to promote the local use of resources \cite{AERtrial}, DNSPs only charge the CBS operator when electricity is imported from the power grid. Therefore, no network charges are applied when CBS is charged using local PV generation. The last term in \eqref{eqn:CBS_profit} represents incentives to reduce the maximum peak demand in the local community, where $\zeta^\text{user}_h \! := \! \max \{\sum_{n \in N} \! \big( x^+_{n,h,t} \! - \! x^-_{n,h,t} \big) | t \! \in \! \mathcal{T} \}$ represents the peak demand of the local end users. Constraints \eqref{eqn:net_demand} and \eqref{eqn:peak_definition} denote the net and peak demand of the entire community with respect to the net demand of the local end users and the CBS (dis)charging power. For compactness, we represent the complementarity constraints as in \eqref{eqn:complementarity_local_net} to avoid $\upsilon^+_{h,t}$ and $\upsilon^-_{h,t}$ simultaneously taking non-zero values. Constraints \eqref{eqn:battery_soc}--\eqref{eqn:soc_boundary} represent the operation and physical conditions of the CBS. In particular, the evolution of the CBS state-of-charge (SoC) is calculated by \eqref{eqn:battery_soc}. Constraint \eqref{eqn:charging_power} defines the CBS (dis)charging power, while \eqref{eqn:charging_limit} and \eqref{eqn:soc_boundary} denote its physical boundaries. Lastly, constraint \eqref{eqn:LUoS} denotes the imported electricity from the utility grid used for CBS charging.

\subsection{Extending the deterministic model to consider bounded rationality via a chance-constrained model}
\label{subsec:CC_model}
As mentioned in subsection \ref{subsec:bounded_rationality}, the CBS operator only receives HEMS price-responsive consumption without considering consumption randomness. This can result in significant financial losses for the CBS operator. To deal with random behaviour of end users, we extend the CBS operation model to a stochastic chance-constrained optimisation problem by incorporating the consumption randomness. Under the RHO regime, the optimisation is solved sequentially with updated information about the end users. This allows the CBS operator to refine its understanding of end-user behaviour based on the observed consumption randomness using the \textit{conditioning} operation of Gaussian processes. For example, consider a receding horizon starting at $h$. We have the observed consumption randomness at $h$ defined as $\mathbf{x}^{\text{rnd}\ast}_n = (x^{\text{rnd}\ast}_{n,1}, \! \cdots \!, x^{\text{rnd}\ast}_{n,h})^\top$, and $\mathbf{\hat{x}}^{\text{rnd}}_n = (\hat{x}^\text{rnd}_{n,h,1}, \! \cdots \!, \hat{x}^\text{rnd}_{n,h,T})^\top$ as the randomness estimation for one day ahead from $h$. We then have a joint distribution:
\begin{flalign}
\begin{bmatrix} 
\mathbf{x}^{\text{rnd}\ast}_n \\
\mathbf{\hat{x}}^\text{rnd}_n
\end{bmatrix}
\sim \mathcal{N}
\left(
\begin{bmatrix} 
\boldsymbol{0} \\
\boldsymbol{0} 
\end{bmatrix},
\begin{bmatrix} 
\mathbf{K}_{\mathcal{X}_n^\ast \mathcal{X}_n^\ast} + \sigma^2_{\varepsilon_n}I & \mathbf{K}_{\mathcal{X}_n^\ast \mathcal{\hat{X}}_n} \\
\mathbf{K}_{\mathcal{\hat{X}}_n \mathcal{X}_n^\ast} & \mathbf{K}_{\mathcal{\hat{X}}_n \mathcal{\hat{X}}_n}
\end{bmatrix}
\right)
\end{flalign}
where $\mathbf{K}_{\mathcal{X}_n^\ast \mathcal{X}_n^\ast} \! = \! k_n(\mathcal{X}_n^\ast \mathcal{X}_n^\ast)$, $\mathbf{K}_{\mathcal{\hat{X}}_n \mathcal{\hat{X}}_n} \! = \! k_n(\mathcal{\hat{X}}_n \mathcal{\hat{X}}_n)$, $\mathbf{K}_{\mathcal{X}_n^\ast \mathcal{\hat{X}}_n} \! = \! k_n(\mathcal{X}_n^\ast \mathcal{\hat{X}}_n) \! = \! \mathbf{K}_{\mathcal{\hat{X}}_n \mathcal{X}_n^\ast}^\top$. The refined understanding for $\mathbf{\hat{x}}^\text{rnd}_n$ is:
\begin{flalign}
\label{eqn:conditional_GP}
    \mathbf{\hat{x}}^\text{rnd}_n | \mathbf{x}^{\text{rnd}\ast}_n  & \! \sim \mathcal{N} \bigg(\! \mathbf{K}_{\mathcal{\hat{X}}_n \mathcal{X}_n^\ast} (\mathbf{K}_{\mathcal{X}_n^\ast \mathcal{X}_n^\ast} + \sigma^2_{\varepsilon_n}I)^{-1} \mathbf{x}^{\text{rnd}\ast}_n, \nonumber \\
    & \mathbf{K}_{\mathcal{\hat{X}}_n \mathcal{\hat{X}}_n} \!\! - \! \mathbf{K}_{\mathcal{\hat{X}}_n \mathcal{X}_n^\ast} \! (\mathbf{K}_{\mathcal{X}_n^\ast \mathcal{X}_n^\ast} \! + \! \sigma^2_{\varepsilon_n}I)^{\! -1} \mathbf{K}_{\mathcal{X}_n^\ast \mathcal{\hat{X}}_n} \!\! \bigg) \!. &&
\end{flalign}

To incorporate the probabilistic estimation of consumption randomness into the CBS operation model, we modify constraints \eqref{eqn:net_demand} and \eqref{eqn:LUoS} as follows:
\begin{subequations}
\label{eqn:probability_constraint}
\begin{flalign}
\label{eqn:net_demand_probability}
    \mathbb{P} \bigg( \! \upsilon^+_{h,t} \! \geq \!\! \sum_{n \in N} \!\! \big( x^+_{n,h,t} \! - \! x^-_{n,h,t} \! + \! \hat{x}^\text{rnd}_{n,h,t} \big) \! + \! P_{h,t} \Delta t \! \! \bigg) \! \geq \eta_1, &&
\end{flalign}
\begin{flalign}
\label{eqn:LUoS_probability}
    \mathbb{P} \bigg( \! \upsilon^{\text{grid}}_{h,t} \geq P^{\text{ch}}_{h,t} \Delta t - \sum_{n \in N} \! \big( x^-_{n,h,t} - \hat{x}^\text{rnd}_{n,h,t} \big) \bigg) \geq \eta_2, &&
\end{flalign}
\end{subequations}
where $\eta$ denotes the tolerance probability required for the constraint inside $\mathbb{P}(\cdot)$. Note that we added the expected randomness consumption $\hat{x}^\text{rnd}_{n,h,t}$ to both constraints. Since $\mathbf{\hat{x}}^\text{rnd}_{n}$ follows a multivariate Gaussian distribution, we leverage the \textit{marginalisation} operation to get the marginalised distribution for each $\hat{x}^\text{rnd}_{n,h,t} \! \sim \! \mathcal{N}(\hat{\mu}^\text{rnd}_{n,h,t}, \hat{\sigma}^{\text{rnd}^2}_{n,h,t} \!)$. To this end, we can analytically formulate the chance constraints in \eqref{eqn:probability_constraint} as follows:
\begin{subequations}
\label{eqn:chance_constraint}
\begin{flalign}
\label{eqn:net_demand_cc}
    \upsilon^+_{h,t} \geq \!\! \sum_{n \in N} \!\! \big( x^+_{n,h,t} \! - \! x^-_{n,h,t} \! \big) + \hat{\mu}^\text{rnd}_{h,t} + \hat{\sigma}^\text{rnd}_{h,t} \Phi^{-1} \! (\eta_1) + P_{h,t} \Delta t, &&
\end{flalign}
\begin{flalign}
\label{eqn:LUoS_cc}
    \upsilon^{\text{grid}}_{h,t} \geq P^{\text{ch}}_{h,t} \Delta t - \!\! \sum_{n \in N} \!\! \big( x^-_{n,h,t}\big) + \hat{\mu}^\text{rnd}_{h,t} + \hat{\sigma}^\text{rnd}_{h,t} \Phi^{-1} \! (\eta_2), &&
\end{flalign}
\end{subequations}
where $\hat{\mu}^\text{rnd}_{h,t} = \sum_{n \in N} \hat{\mu}^\text{rnd}_{n,h,t}$, $\hat{\sigma}^\text{rnd}_{h,t} = \sqrt{\sum_{n \in N} \hat{\sigma}^{\text{rnd}^2}_{n,h,t}}$, and $\Phi^{-1}(\cdot)$ denotes the inverse cumulative distribution function of the Gaussian distribution. Note that the reformulation of \eqref{eqn:net_demand_cc} removes the variable $\upsilon^-_{h,t}$. Therefore, in the stochastic optimisation model, we remove the constraint \eqref{eqn:complementarity_local_net} and replace it with a new constraint to avoid $\upsilon^+_{h,t}$ increasing indefinitely. The new constraint is as follows:
\begin{flalign}
\label{eqn:restrict_local_net_positive}
    \upsilon^+_{h,t} \leq \! \sum_{n \in N} \!\! \big( x^+_{n,h,t} \big) + \hat{\mu}^\text{rnd}_{h,t} + \hat{\sigma}^\text{rnd}_{h,t} \Phi^{-1} \! (\eta_3) + P_{h,t} \Delta t, &&
\end{flalign}

\subsection{Revenue calculation}
\label{subsec:revenue_calc}

In \eqref{eqn:CBS_profit}, we formulate the net operating cost as the objective function for each receding horizon. In particular, the given cost depends on the incentives provided for peak demand reduction. However, in reality, peak demand reduction is not assessed on a daily basis but over a billing period, typically a year. For this reason, after obtaining the results from the CBS optimisation model, we recalculate the profit of the CBS operator as:
\begin{flalign}
\label{eqn:final_revenue}
    \text{Profit} = \frac{T^\text{e}}{H} & \bigg[ -\!\! \sum_{h \in \mathcal{H}} \!\! \big[ \lambda^\text{RT}_h (\upsilon^{+\ast}_h - x^{\text{grid}\ast}_{n,h}) + \lambda^{\text{grid}} \upsilon^{\text{grid}\ast}_h \nonumber \\
    & + \lambda^{\text{Opex}} P^{\text{ch}\ast}_h \Delta t \big] +  \lambda^\text{peak} \zeta^\text{peak} \! - \! \lambda^\text{fix} E^\text{cap} C^\text{rate}  \nonumber \\ 
    & + \!\! \sum_{n \in N} \!\! \big(\lambda^\text{cre} \delta^\ast_{n, H} \! - \! \lambda^\text{rem} C^\ast_{n,H}\big) \! \bigg] \! - \! E^\text{cap} E^\text{cost}, &&
\end{flalign}
where $\zeta^\text{peak} = \max_{h \in \mathcal{H}} \{\sum_{n \in N} \! \big( x^{+\ast}_{n,h} + x^\text{rnd}_{n,h} \! - \! x^{\text{Exp}\ast}_{n,h} \big)\} - \max_{h \in \mathcal{H}} \{\sum_{n \in N} \! x^\text{rnd}_{n,h} + \upsilon^{+\ast}_h \}$. In addition to the financial terms outlined in \eqref{eqn:CBS_profit}, we consider realistic fixed costs and fixed revenues to estimate the realistic profit made by the CBS operator. They include the electricity payment from the end users, $\lambda^\text{RT}_h\! x^{\text{grid}\ast}_{n,h}$, a fixed network charge, $\lambda^\text{fix} E^\text{cap} C^\text{rate}$, paid to the DNSP based on the CBS power capacity \cite{AusrgidCBStariff}, and solar credit usage charge paid by solar prosumers, $\lambda^\text{cre} \delta^\ast_{n, H}$, for the CBS usage. To ensure the highest benefits for the prosumers, the CBS operator also pays a fixed rate for the remaining solar credits, $\lambda^\text{rem} C^\ast_{n,H}$, at the end of the billing period. Lastly, we scale the calculation by CBS lifetime and subtract the battery cost $E^\text{cap} E^\text{cost}$.

\section{Simulation study}
\label{sec:sim_study}

In this section, we analyse the impact of end-user irrationality and evaluate the financial benefits of our chance-constrained CBS operation model. We run the simulation based on real-world data and compare the end-user benefits under the proposed scheme against another existing retailer.

\subsection{Simulation setup}
\subsubsection{End-user profiles} 
We randomly selected 50 end users' profiles from the 2012 Solar Home dataset in New South Wales (NSW), Australia \cite{Ausgrid2012data}. These profiles were equally divided into two groups: 25 solar prosumers with rooftop solar PV systems and 25 non-solar consumers. Due to the increase in rooftop PV capacity, the PV generation profiles were uniformly scaled up three times, resulting in an average PV capacity of 5.5 kWp. For the simulation period, we chose the first week of each month (84 days in total) to capture the seasonality and variability of the end-user behaviour and PV generation throughout the year. In this study, we assumed a billing period that lasted for a season, while the reduction in peak demand was evaluated at the end of the year. 

\subsubsection{Electricity prices and charges} 
Similar to the simulation period, we picked the RT wholesale market prices in NSW for the first week of each month in 2021 \cite{AEMOprice}. The network charges for end users and CBS operator were taken from the DNSP in NSW, i.e., AusGrid \cite{Ausgridtariffs}. In particular, end users were subject to tariff codes EA959/EA960 while the CBS operator was assigned tariff codes EA962/EA963 \cite{AusrgidCBStariff}. Although the peak demand reduction service is not currently available, we hypothesised that DNSP would incentivise the CBS operator at the peak demand tariff rate EA302 \cite{Ausgridtariffs}, traditionally charged to business customers.

\subsubsection{CBS data and other parameters} 
In this study, we refer to the CBS specifications outlined in \cite{batterycostCSIRO}, and summarise the CBS and other simulation parameters in Table \ref{tab:simu_input}. All monetary values are in Australian dollars (AUD). As the price elasticity of the end users can vary throughout the day \cite{price2020zsuzanna}, we adopted three-time bands and uniformly distributed the coefficients of elasticity among the end users. We represent in Table \ref{voltable} the distribution ranges, where the time band is defined by the NSW DNSP \cite{Ausgridtariffs}. To represent the diversity of end users in the community, we also uniformly distributed the discounting degree, $\kappa_n$, with the specified range presented in Table \ref{tab:simu_input}.

\begin{table}[t]
  \caption{CBS specifications and simulation parameters}
  \label{tab:simu_input}
  \centering
  \begin{tabular}{p{0.07\textwidth}r|p{0.075\textwidth}r}
  \specialrule{.15em}{0em}{0.2em} \multicolumn{2}{l}{\textbf{CBS specifications}} & \multicolumn{2}{l}{\textbf{Simulation parameters}} \\
  \specialrule{.1em}{0em}{0.2em}
  $\Gamma$ & $90$\% & $\underline{x}_{n,t}$, $\overline{x}_{n,t}$ & $0.5\hat{x}_{n,t}$, $1.5\hat{x}_{n,t}$ \\
  $\underline{\text{SoC}}$, $\overline{\text{SoC}}$ & $0$\%, $100$\% & $T$ & $48$ (24 hours) \\
  $E^\text{cap}$ & $100$ kWh & $T^\text{RB}$ & $12$ (6 hours) \\
  $C^\text{rate}$ & $0.5$ & $\tau$ & $0.2$ \\
  $T^e$ & $3650$ (10 years) & $\kappa_n$ & $[0.1, \, 0.5]$ \\
  $E^\text{cost}$ & AUD\$ 800/kWh & $\sigma_\text{SE}$ & $1$ \\
  $\lambda^\text{Opex}$ & $2.2$ \textcent/kWh & $\eta_1, \eta_2, \eta_3$ & $99.9 \%, 97.5 \%, 99.9 \%$ \\
  $\Delta t$ & $0.5$ hours & $\lambda^\text{cre}, \lambda^\text{rem}$ & $10$ \textcent/kWh, $5$ \textcent/kWh \\
  \specialrule{.1em}{0.1em}{0em}
  \end{tabular}
\end{table}

\begin{table}[!t]
    \caption{Time-varying price elasticity coefficients range}
    \centering
    \label{voltable}
    \begin{tabular}{c|c|c|c}
    \toprule
    \rowcolor{GrayTop} & \textbf{Off-peak} & \textbf{Shoulder}& \textbf{Peak} \\
    \midrule
    \textbf{Price elasticity} & $[-0.2, \; -0.3]$ & $[-0.3, \; -0.5]$ & $[-0.5, \; -0.7]$ \\
    \bottomrule
    \end{tabular}
\end{table}

\subsection{Simulation results}
\subsubsection{Local price-responsive consumption}

\begin{figure}[!t]
    \centering
    \includegraphics[trim={0.25cm 0.3cm 0.21cm 0.21cm}, clip, width=\linewidth]{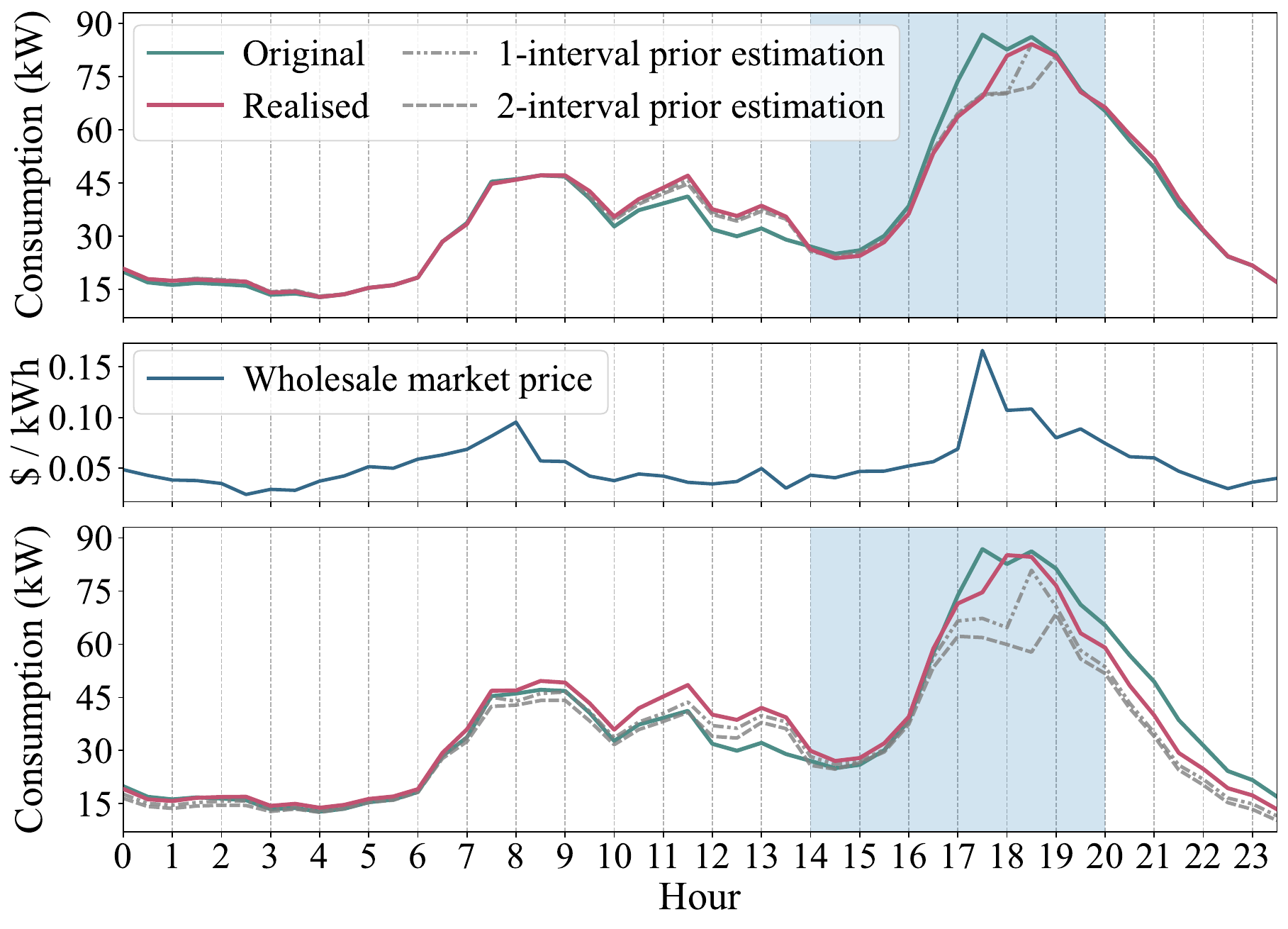}
    \caption{Comparison of local consumption of a winter weekday. The top figure depicts consumption considering only loss aversion (\textbf{LA}), while the bottom figure includes both loss aversion and time inconsistency properties (\textbf{LA+TI})}
    \label{fig:local_consumption}
\end{figure}

Figure \ref{fig:local_consumption} shows the aggregated price-responsive consumption of the local community on a weekday in winter obtained from the end-user optimisation model in subsections \ref{subsec:loss_aversion} and \ref{subsec:time_inconsis}. 
The pink lines represent the realised price-responsive consumption, while the dashed grey lines represent the expected consumption estimated one and two intervals prior. For example, the value indicated by the first grey line at 15:00 signifies that consumption was estimated at the receding horizon starting at 14:30. Meanwhile, the second grey line's value at 15:00 represents the consumption estimation made at the receding horizon starting at 14:00. The bottom figure compared to the top one exhibits fewer overlaps among the three lines, demonstrating the changing decisions of end users over time. In this way, we eliminate the presumption of a static and perfect consumption forecast by end users through time inconsistency. This aligns closely with real-world scenarios where the accuracy of consumption forecasts, such as those generated by ML models, varies based on the lookahead horizon \cite{probabilistic2021julian}. The separation of the three lines in the top figure at 18:00--19:00 comes from the updated information as time proceeds and the short-horizon rebound effect. This rebound effect also explains why consumption remains relatively high during peak hours, which charges 27\textcent/kWh, as defined by DNSP \cite{AusrgidCBStariff}. Peak hours specified by the DNSP are depicted by the blue-shaded region in Fig. \ref{fig:local_consumption}. Although consumption sees an overall reduction during peak hours compared to original values (green lines) due to the price-responsive behaviour implemented in \eqref{eqn:time_incons}, maximum peak power demand remains high. In particular, the peak is shifted from 17:30 to 18:00--18:30 due to the high RT price at 17:30, as shown in the middle figure. This underscores the necessity of CBS for reducing peak demand, as will be discussed later in subsection \ref{sec:CBS_oper_profit}.

\subsubsection{Randomness of consumption}
Even with the help of HEMS, end users can still deviate significantly due to their random behaviour. As outlined in subsection \ref{subsec:bounded_rationality}, after extracting the randomness behaviour using MSTL decomposition, we apply an outlier filtering and perform a normality check. Table \ref{tab:normality_check} shows the average percentage of outliers removed by the IQR thresholding strategy and the intervals that passed the KS normality test in all end-user randomness time series. We incorporate the eliminated outliers into the Gaussian noise term $\varepsilon_{n,h}$ by setting $\sigma_{\varepsilon_{n,h}} = 0.1 \, \sigma_{n,h}$.

\begin{table}[!t]
    \caption{Average percentage of removed outliers and the intervals that passed normality test}
    \centering
    \label{tab:normality_check}
    \begin{tabular}{c|c|c|c|c}
    \toprule
    \rowcolor{GrayTop} & \textbf{Spring} & \textbf{Summer} & \textbf{Autumn} & \textbf{Winter} \\
    \midrule
    \textbf{Outliers} & 5.0\% & 5.2\% & 5.0\% & 5.4\% \\
    \textbf{Normality} & 99.8\% & 99.6\% & 99.6\% & 99.7\% \\
    \bottomrule
    \end{tabular}
\end{table}

\begin{figure}[!t]
    \centering
    \includegraphics[trim={0.25cm 0.3cm 0.21cm 0.21cm}, clip, width=0.93\linewidth]{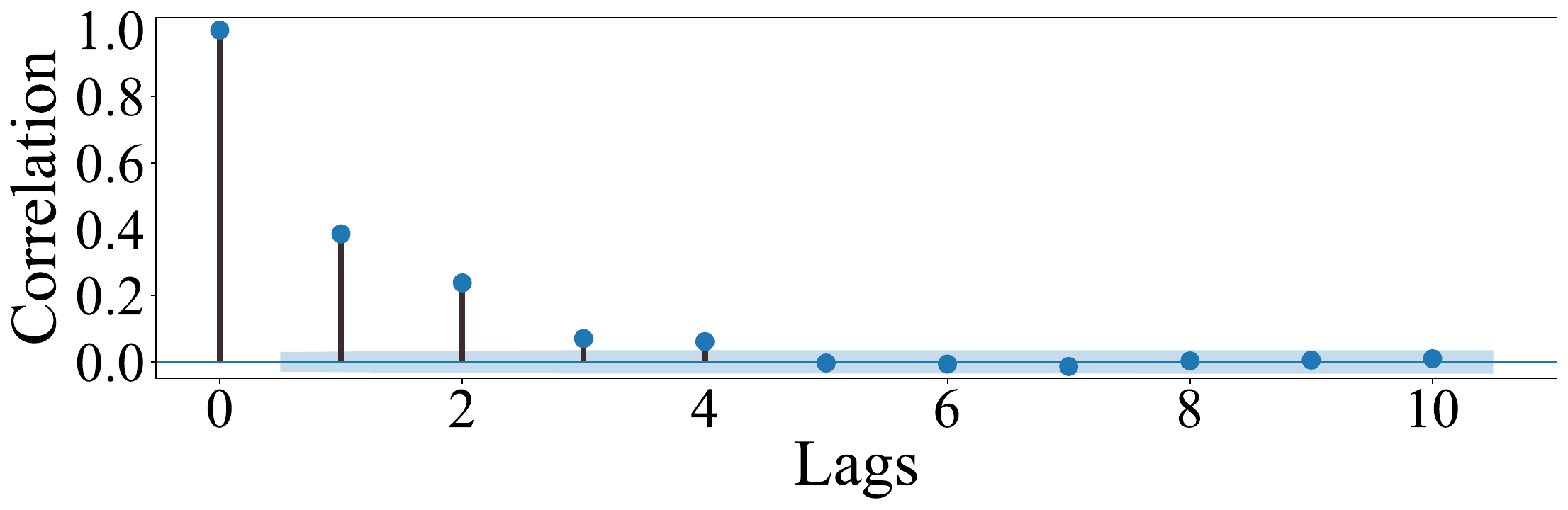}
    \caption{The ACF plot of an end user's randomness time series}
    \label{fig:acf_plot}
\end{figure}

\begin{figure}[!t]
    \centering
    \includegraphics[trim={0.25cm 0.3cm 0.21cm 0.21cm}, clip, width=0.68\linewidth]{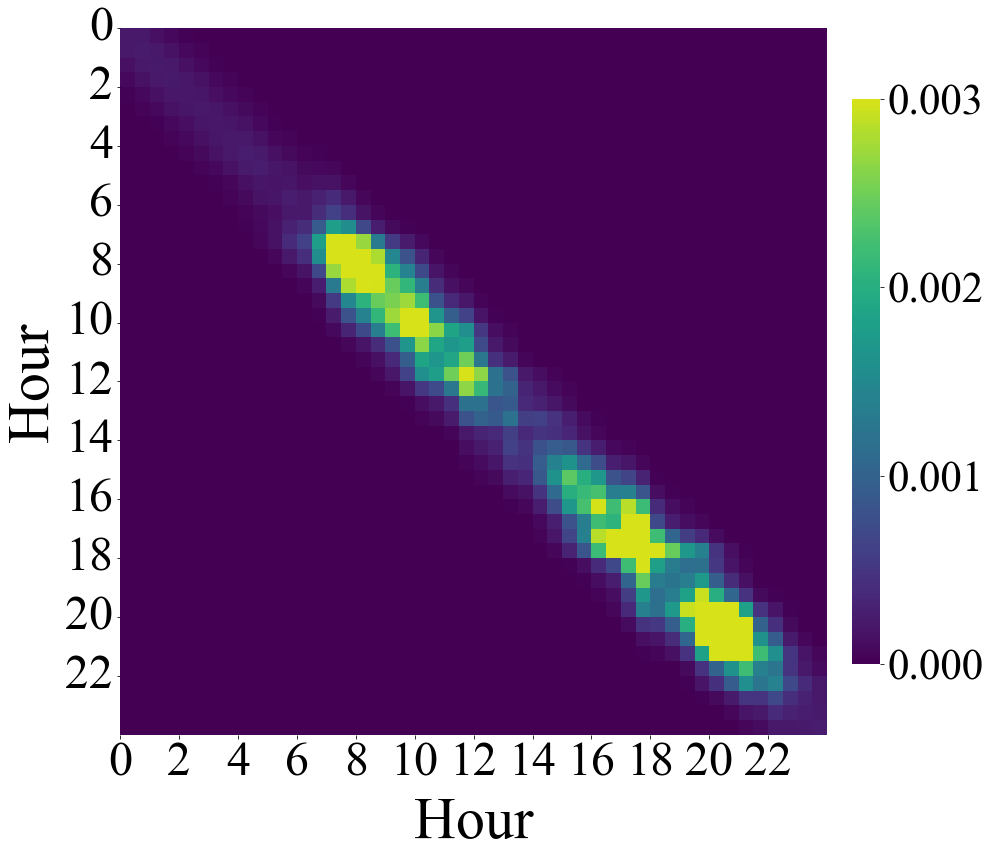}
    \caption{Randomness covariance matrix of an end user generated from the proposed non-stationary kernel}
    \label{fig:cov_matrix}
\end{figure}

Next, an autocorrelation check is performed on the filtered randomness time series. Figure \ref{fig:acf_plot} shows the ACF plot for a representative end user. Based on the identified average significant lags across all end users, we assign a value of $2.1$ to $l_\text{SE}$ in \eqref{eqn:SE_kernel}. This parameter can be interpreted as defining the autocorrelation lag order for the samples generated from the Gaussian processes. Specifically, a larger $l_\text{SE}$ value increases the influence of neighbouring half-hourly intervals, implying a higher autocorrelation order in consumption time series. While it is best to assign unique $l_\text{SE}$ values for each end user based on their specific lag order, for simplicity, we opt for a uniform value of $l_\text{SE} = 2.1$ among all users. Figure \ref{fig:cov_matrix} shows the randomness covariance matrix generated from the proposed non-stationary kernel of the same end user as in Fig. \ref{fig:acf_plot}. For each specific interval, we can see the influence of the surrounding intervals on the randomness value at that interval based on the chosen value of $l_\text{SE}$. Note that given the \textit{causal} nature of the model, the randomness value at interval $t$ is only affected by the preceding randomness values. Unlike stationary kernels, the values along the diagonal here are non-uniform \cite{advanced2022marcus}, offering greater flexibility to capture varying randomness at different times of day. The randomness covariance matrix shows that the end user on average has a high randomness measure early in the morning and during the evening peak hours, aligned with periods of increased home activity. 

\begin{figure}[!t]
    \centering
    \includegraphics[trim={0.25cm 0.5cm 0.21cm 0.21cm}, clip, width=0.98\linewidth]{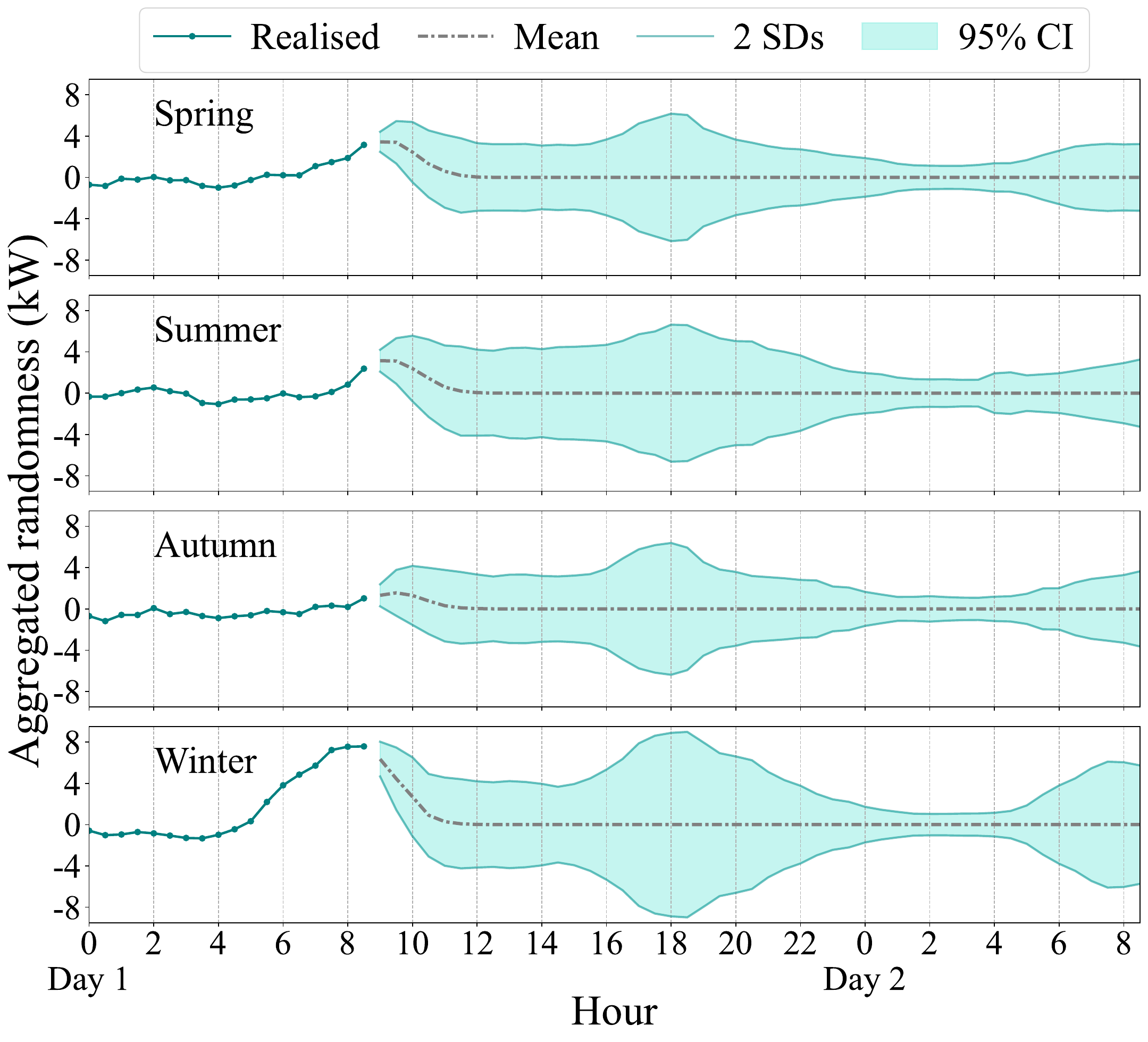}
    \caption{Aggregated randomness for different seasons}
    \label{fig:agg_randomness}
\end{figure}

To demonstrate the impact of seasonality, Fig. \ref{fig:agg_randomness} presents the aggregated randomness of the local community throughout all four seasons with the receding horizon starting at 9:00 on day 1. The 95\% confidence interval (CI) for the day ahead is constructed using two standard deviations (SDs) derived from the diagonal of the randomness covariance matrix aggregated from all end users. In addition to the time-of-day-dependent randomness explained above, we observe in the figures that the distribution is tightened at the start of the receding horizon, i.e., at 9:00, reflecting the conditioning operation of the Gaussian processes. This illustrates how the CBS operator can refine its expectations of future randomness based on previously observed randomness values. Such a mechanism of dynamically updating expectations is consistent with real-world scenarios, where the accuracy of forecasts generally improves as they approach the actual time of occurrence \cite{probabilistic2021julian}. Lastly, it can be seen that the randomness measure is highest during winter mainly due to increased consumption. In general, the results show that as consumption increases, the unpredictability of consumption patterns also increases. 

The final consumption pattern of each end user in our model is derived by combining the three behavioural economic properties. This involves summing the randomness component from bounded rationality (\textbf{BR}), as depicted in Fig. \ref{fig:agg_randomness}, with the consumption pattern influenced by loss aversion and time inconsistency (\textbf{LA+TI}), as illustrated at the bottom of Fig. \ref{fig:local_consumption}. Since the randomness is zero reverting, the model retains the price-responsive behaviour, which is modelled in the \textbf{LA+TI} properties. However, in the short term, this pattern is overlaid with noise due to the inherent randomness of the end users, reflecting the complex and often unpredictable nature of end-user behaviour. Intuitively, compared to ML forecasting models, which rely on exogenous variables, the results produced from \textbf{LA+TI} are explainable by variables such as temporal information, market prices, and demographic factors. Specifically, time inconsistency parameters correlate with the age of the occupants, while price elasticity parameters can be adjusted according to household income levels \cite{price2020zsuzanna}. Meanwhile, \textbf{BR} concentrates on the internal dynamics of consumption time series, encapsulating the autocorrelation (lag values) inherent in the consumption data. This means that it considers how past consumption values influence future usage, a key factor in accurately predicting consumption patterns \cite{probabilistic2021julian}. Additionally, \textbf{BR} captures the residual or randomness elements that are inherent in ML forecasting models. These elements refer to the unpredictable fluctuations in consumption that cannot be explained solely by exogenous variables or past trends.

\subsubsection{CBS operator profit}
\label{sec:CBS_oper_profit}

\begin{figure}[!t]
    \centering
    \includegraphics[trim={0.25cm 2.5cm 0.21cm 23.5cm}, clip, width=\linewidth]{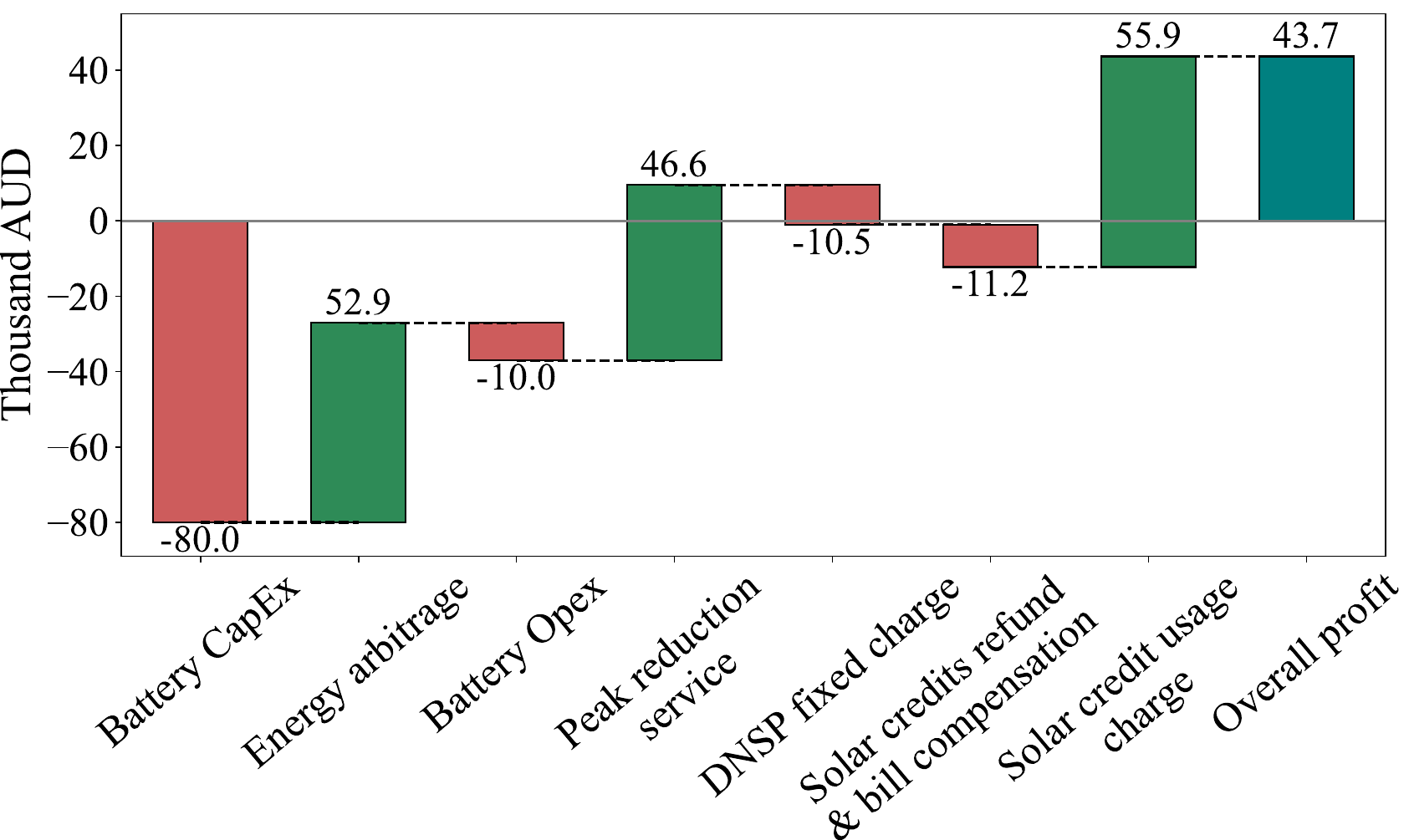}
    \caption{CBS operator's profit breakdown of the proposed model. Battery cost is estimated at \$800/kWh, and the solar credit usage charge is set at 10\textcent/kWh}
    \label{fig:profit_breakdown}
\end{figure}

\begin{figure}[!t]
    \centering
    \includegraphics[trim={0.25cm 0.3cm 0.31cm 0.21cm}, clip, width=0.74\linewidth]{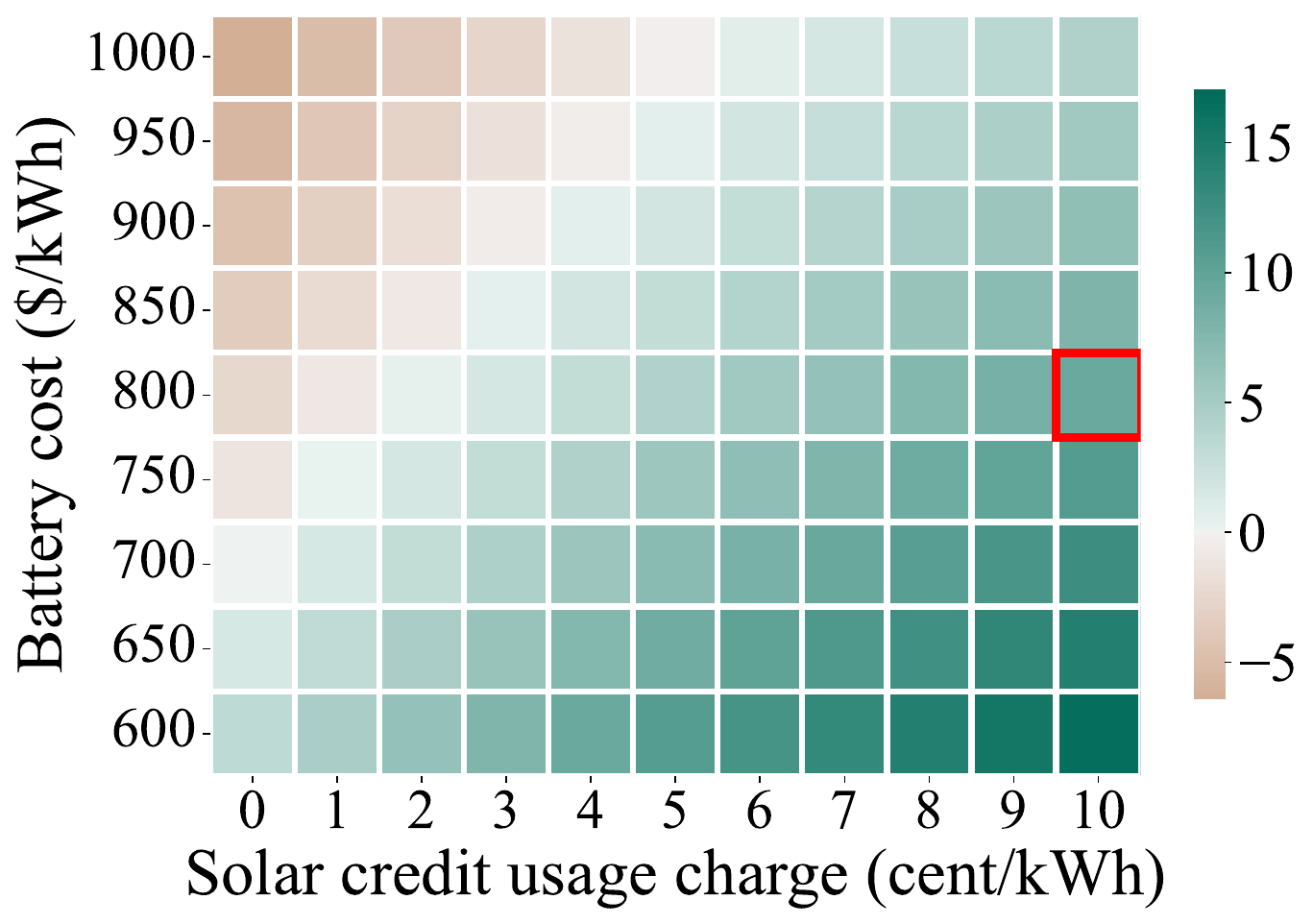}
    \caption{Project IRR (in \%) with respect to battery cost and solar credit usage charge}
    \label{fig:IRR_values}
\end{figure}

\begin{table}[!t]
    \caption{Annual revenue from peak reduction and energy arbitrage under three different scenarios of CBS operation}
    \centering
    \label{tab:peak_reduction}
    \begin{tabular}{c|c|c|c|c}
    \toprule
    \rowcolor{GrayTop} \begin{tabular}{@{}c@{}}\textbf{Irrationality}\\ \textbf{properties}\end{tabular} & \begin{tabular}{@{}c@{}}\textbf{Peak} \\ \textbf{reduction}\end{tabular} & \begin{tabular}{@{}c@{}}\textbf{Peak} \\ \textbf{service}\end{tabular} & \begin{tabular}{@{}c@{}}\textbf{Energy} \\ \textbf{arbitrage}\end{tabular} & \begin{tabular}{@{}c@{}}\textbf{Annual} \\ \textbf{revenue}\end{tabular} \\
    \midrule
    \textbf{LA (existing)} & 27.96 kW & \$3.18k & \$3.94k & \$7.12k \\
    \textbf{LA + TI} & 35.21 kW & \$4.34k & \$3.96k & \$8.30k \\
    \textbf{LA + TI + BR} & \textbf{37.82 kW} & \textbf{\$4.66k} & \textbf{\$5.29k} & \textbf{\$9.95k} \\
    \bottomrule
    \end{tabular}
\end{table}

Using the price-responsive consumption reported by the end users, as shown in Fig. \ref{fig:local_consumption}, the CBS operator can solve a deterministic optimisation problem to operate the CBS. However, recognising that the HEMS might overlook the randomness behaviour, the CBS operator can instead opt for a chance-constrained operational model. This model considers the additional distribution shown in Fig. \ref{fig:agg_randomness}. Table \ref{tab:peak_reduction} shows the annual revenue achieved by CBS in three different scenarios: first, the CBS operator only considers \textbf{LA} when solving for CBS operation, corresponding to existing work in the literature \cite{optimal2022Dinh,accomodating2019radoszynski}; second, both \textbf{LA+TI} are considered; third, all three properties of irrationality are considered (\textbf{LA+TI+BR}) using chance-constrained optimisation. Note that in this analysis, end users display all three behavioural properties in all scenarios. The results show that the third scenario delivers the highest peak demand reduction of 37.82 kW, which is approximately \$46.6k of revenue in the peak reduction service. Additionally, when considering all three irrationality properties, the CBS operator achieves the highest revenue in energy arbitrage; hence, the highest revenue overall. In particular, the proposed \textbf{LA+TI+BR} case yields 39.7\% and 19.9\% higher revenue than the \textbf{LA} case and \textbf{LA+TI} cases, respectively.

In Fig. \ref{fig:profit_breakdown}, we show the breakdown of the CBS operator's profits over the CBS lifetime for the third scenario. In addition to revenues from energy arbitrage and peak reduction service, battery capital expenditure (CapEx) and solar credit usage charges greatly impact the overall project profit. Here, the battery cost is estimated at \$800/kWh for a 2-hour battery, and the solar credit usage charge is set to 10\textcent/kWh. In Fig. \ref{fig:IRR_values}, we show the internal rate of return (IRR) of the project as a percentage with respect to varying battery costs and solar credit usage charges. The red box indicates IRR (9.28\%) under the same setup as in Fig. \ref{fig:profit_breakdown}.

\begin{figure}[!t]
    \centering
    \includegraphics[trim={0.25cm 0.3cm 0.21cm 0.21cm}, clip, width=0.94\linewidth]{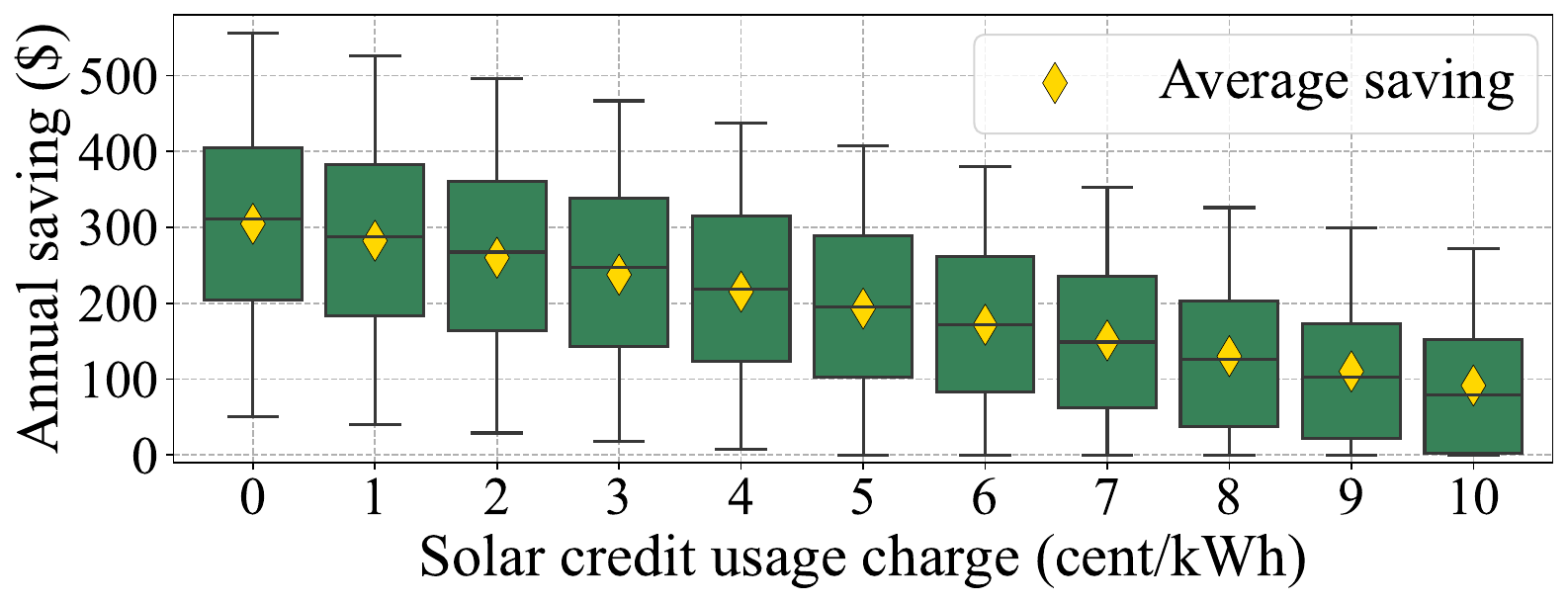}
    \caption{Distribution of solar prosumers annual savings under different solar credit usage charges}
    \label{fig:varying_charge_prosumers}
\end{figure}

\begin{figure}[!t]
    \centering
    \includegraphics[trim={0.25cm 0.3cm 0.21cm 0.21cm}, clip, width=0.98\linewidth]{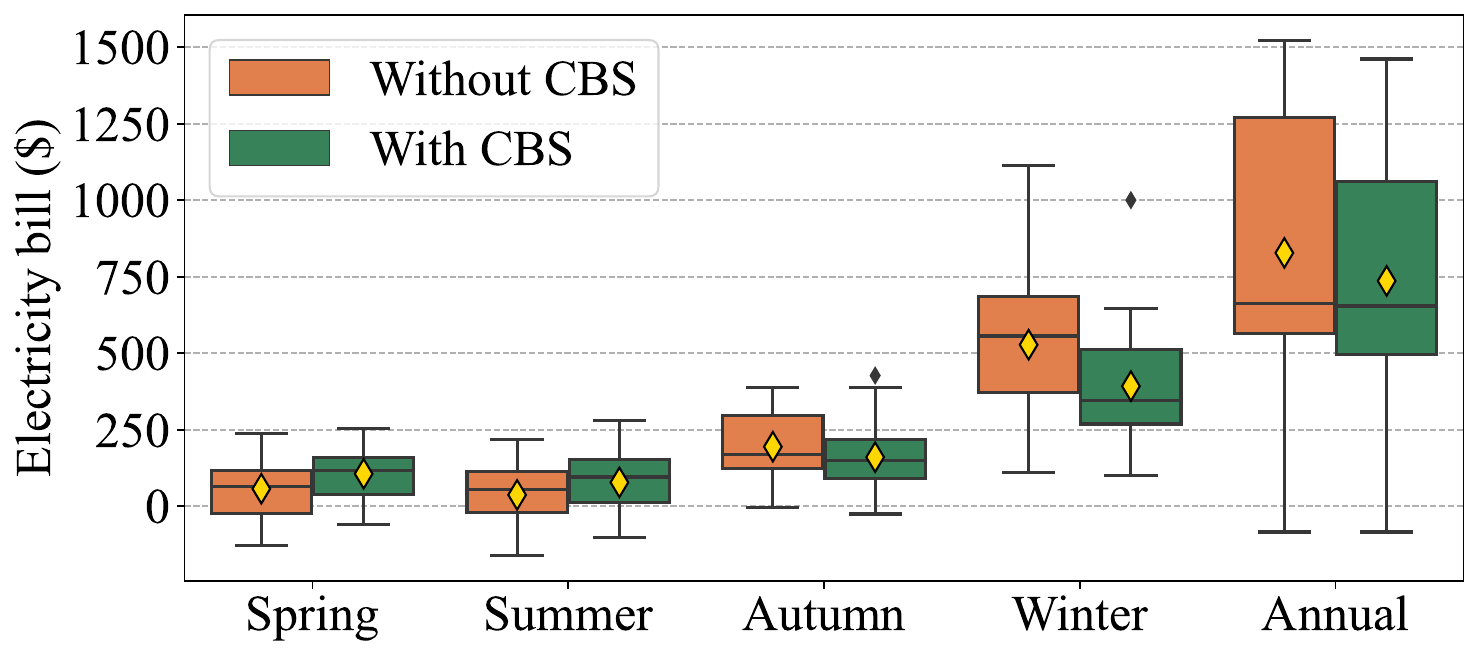}
    \caption{Comparison of solar prosumers' electricity bills when solar credit usage charge is set at 10\textcent/kWh}
    \label{fig:payment_compa}
\end{figure}

\subsubsection{Solar prosumers electricity payment} 
While a high solar credit usage charge yields a higher profit for the CBS operator, it also results in an increase in electricity bill for solar prosumers. In this paper, we compare the electricity bills of prosumers under our proposed scheme with those of Amber Electric \cite{amberelectric}, an existing retailer that sells electricity at wholesale RT prices to residential customers. Figure \ref{fig:varying_charge_prosumers} demonstrates the annual savings of solar prosumers under different credit usage charges. To ensure the highest benefits for the prosumers, a Bill Guarantee scheme is considered to compensate the prosumers if they pay more under our proposed model. For instance, at 10\textcent/kWh credit usage charge, six prosumers might incur a loss under our proposed scheme. Therefore, their savings are zero. Bill compensation is also factored into the CBS operator's profit, as shown in the penultimate column in Fig. \ref{fig:profit_breakdown}. Overall, depending on the strategy of the CBS operator, it can either choose to increase its profit per customer or attract more customers by lowering the credit usage charge.

In Fig. \ref{fig:payment_compa}, we show the distribution of the electricity bills of prosumers under Amber Electric (without CBS) and our proposed scheme (with CBS) at 10\textcent/kWh credit usage charge. Although prosumers on average have lower annual electricity bills (11.1\% reduction), electricity bills during spring and summer are observed to be higher under our proposed model. The exception arises because solar energy is excessive during these seasons and prosumers under Amber Electric can directly sell to the grid at RT prices instead of virtually storing in the CBS. Although in our model, the CBS operator offers a refund at a fixed rate for any remaining solar credits at the end of the billing period, as discussed in subsection \ref{subsec:revenue_calc}, the high RT prices make Amber Electric a slightly more cost-effective option for these two seasons. However, with a higher penetration of RES, more negative prices are expected midday \cite{negativepricing}. As a result, this will make CBS a more lucrative solution for all seasons throughout the year.

\section{Conclusion}
\label{sec:conc}

In this paper, we introduced a price-responsive model of residential users considering their irrational behaviour, encapsulating loss aversion, time inconsistency, and bounded rationality. We incorporated the first two properties into the utility function of the end users, whereas the bounded rationality was modelled through the randomness of the end-user behaviour. To this end, we developed a framework that extracts this random behaviour from the consumption patterns using MSTL and subsequently models this behaviour using non-stationary Gaussian processes. The impact of irrational behaviour is analysed through a business model for a local energy community that integrates CBS. To deal with end users' irrationality, we applied a chance-constrained optimisation to operate the CBS under the RHO regime. Using real-world data from 50 end users, we demonstrated that the proposed DR model incorporating irrationality provides a more realistic estimation of end-user price responsiveness. Next, we quantified the financial benefits of the proposed chance-constrained CBS operation model, resulting in up to 19\% increase in CBS revenue. Also, we showed that solar prosumers under our proposed scheme could have lower electricity bills than existing retailers while still ensuring a profitable business model for the CBS operator. Future work could focus on increasing the CBS operator's profit by providing ancillary services for the electricity market.

\section*{Acknowledgment}
This project is funded jointly by the University of Adelaide industry-PhD grant scheme and Watts A/S, Denmark.




\bibliographystyle{IEEEtran}
\bibliography{reference.bib}

%


\vskip 0pt plus -1fil

\begin{IEEEbiography}[{\includegraphics[width=1in,height=1.25in,clip,keepaspectratio]{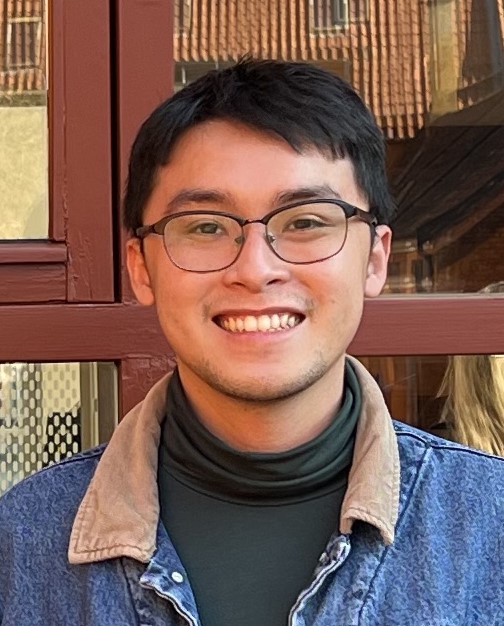}}]{Nam Trong Dinh}
received his BSc (Honours) degree in Electrical and Electronic Engineering from the University of Adelaide, Adelaide, Australia in 2021. He is currently a PhD candidate in the School of Electrical and Mechanical Engineering at the same institution, where he is supported by a joint scholarship funded in collaboration with Watts A/S, Denmark. Also, he is working as a research engineer at Watts A/S as part of the company’s R\&D section. His research interests include demand response, electricity prices forecasting, business models for electricity retailers and aggregators, and battery energy management systems.
\end{IEEEbiography}

\vskip 0pt plus -1fil

\begin{IEEEbiography}[{\includegraphics[width=1in,height=1.25in,clip,keepaspectratio]{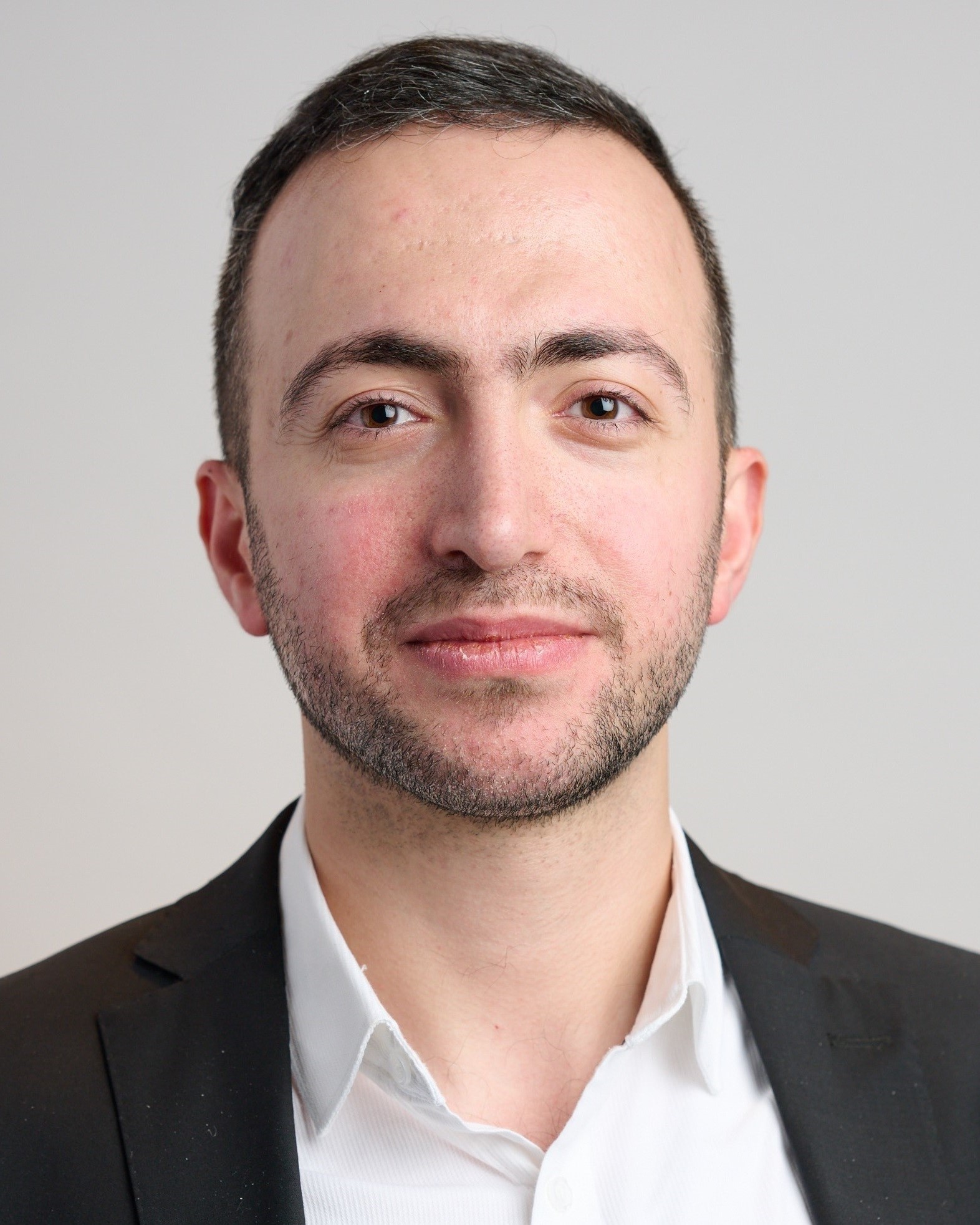}}]{Sahand Karimi-Arpanahi}
received the BSc and MSc degrees in Electrical Engineering from the Sharif University of Technology, Tehran, Iran, in 2016 and 2018, respectively. He worked as a research assistant at Sharif University of Technology from 2018 to 2020. Since 2022, he has been with the Australian Energy Market Operator (AEMO). Also, he is currently a PhD candidate in the School of Electrical and Mechanical Engineering at the University of Adelaide, Adelaide, Australia, and CSIRO Energy, Newcastle, Australia. His research interests include power system optimisation; data analysis in power systems and electricity markets; and grid integration of renewable energy and battery storage systems.
\end{IEEEbiography}

\vskip 0pt plus -1fil

\begin{IEEEbiography}[{\includegraphics[width=1in,height=1.25in,clip,keepaspectratio]{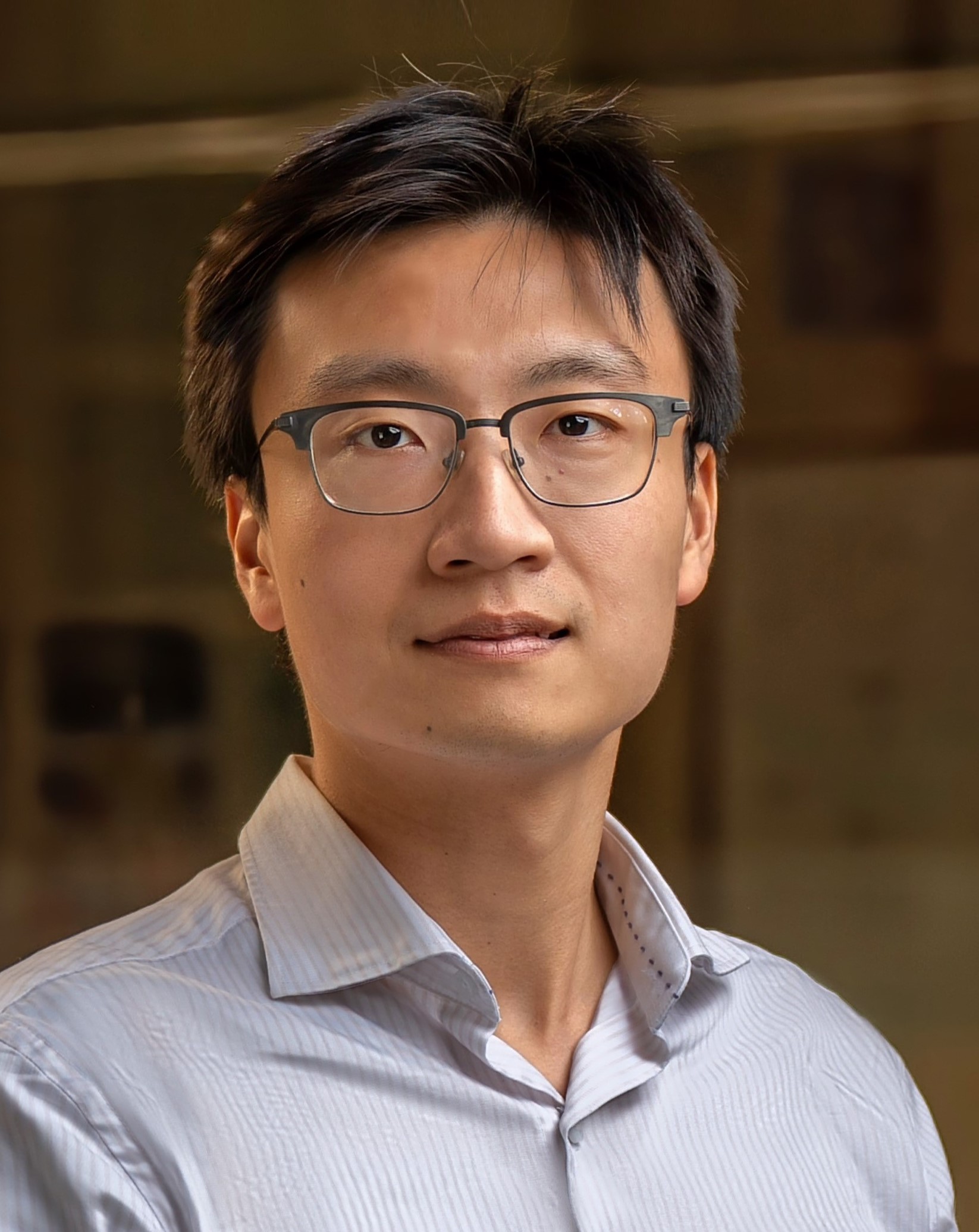}}]{Rui Yuan}
received the B.Sc. degree in telecommunication engineering from the Harbin Institute of Technology, China, in 2016 and the M.Sc. degree in electrical engineering from the University of Melbourne, Australia. He is currently an industry PhD candidate at the University of Adelaide, Australia and a data scientist in Watts A/S, Denmark. His main research interest is analysis and modeling of energy consumption profiles of consumers. This includes data mining, explainable machine learning, synthetic data generation and time series analysis.
\end{IEEEbiography}

\vskip 0pt plus -1fil

\begin{IEEEbiography}[{\includegraphics[width=1in,height=1.25in,clip,keepaspectratio]{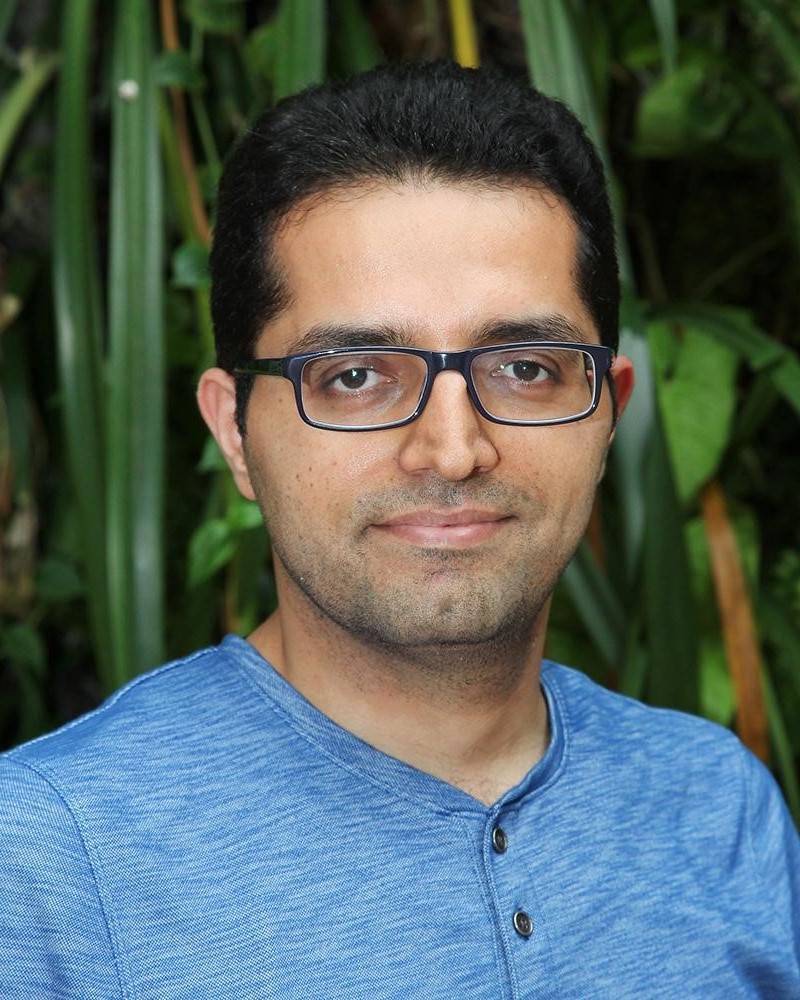}}]{S. Ali Pourmousavi}
(Senior Member, IEEE) received the B.Sc., M.Sc., and Ph.D. degrees (with hons.) in electrical engineering, in 2005, 2008, and 2014, respectively. From 2014 to 2019, he was with California ISO, NEC Laboratories America Inc., Technical University of Denmark, Kongens Lyngby, Denmark, and The University of Queensland, Brisbane, Australia. He is currently a Senior Lecturer in the school of Electrical and Mechanical Engineering, the University of Adelaide, Australia. His current research interests include mining and transportation electrification, BTM flexibility aggregation and demand response.
\end{IEEEbiography}

\vskip 0pt plus -1fil

\begin{IEEEbiography}[{\includegraphics[width=1in,height=1.25in,clip,keepaspectratio]{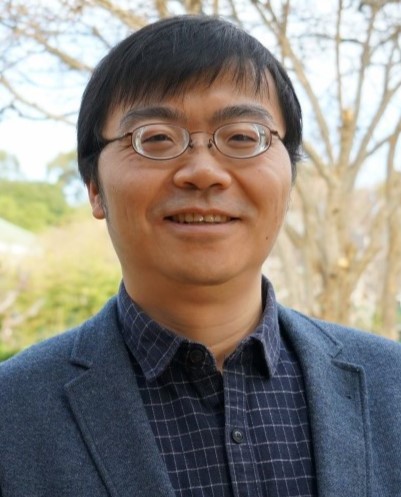}}]{Mingyu Guo}
is a Lecturer in the Optimisation and Logistics group at the University of Adelaide. He received his Ph.D. degree in Computer Science from Duke University, USA. Prior to joining the University of Adelaide, he was a Lecturer in the Economics and Computation group at University of Liverpool, UK. His main research focus is algorithmic game theory and its applications, as well as combinatorial optimisation via neural networks and evolutionary computation. 
\end{IEEEbiography}

\vskip 0pt plus -1fil

\begin{IEEEbiography}[{\includegraphics[width=1in,height=1.25in,clip,keepaspectratio]{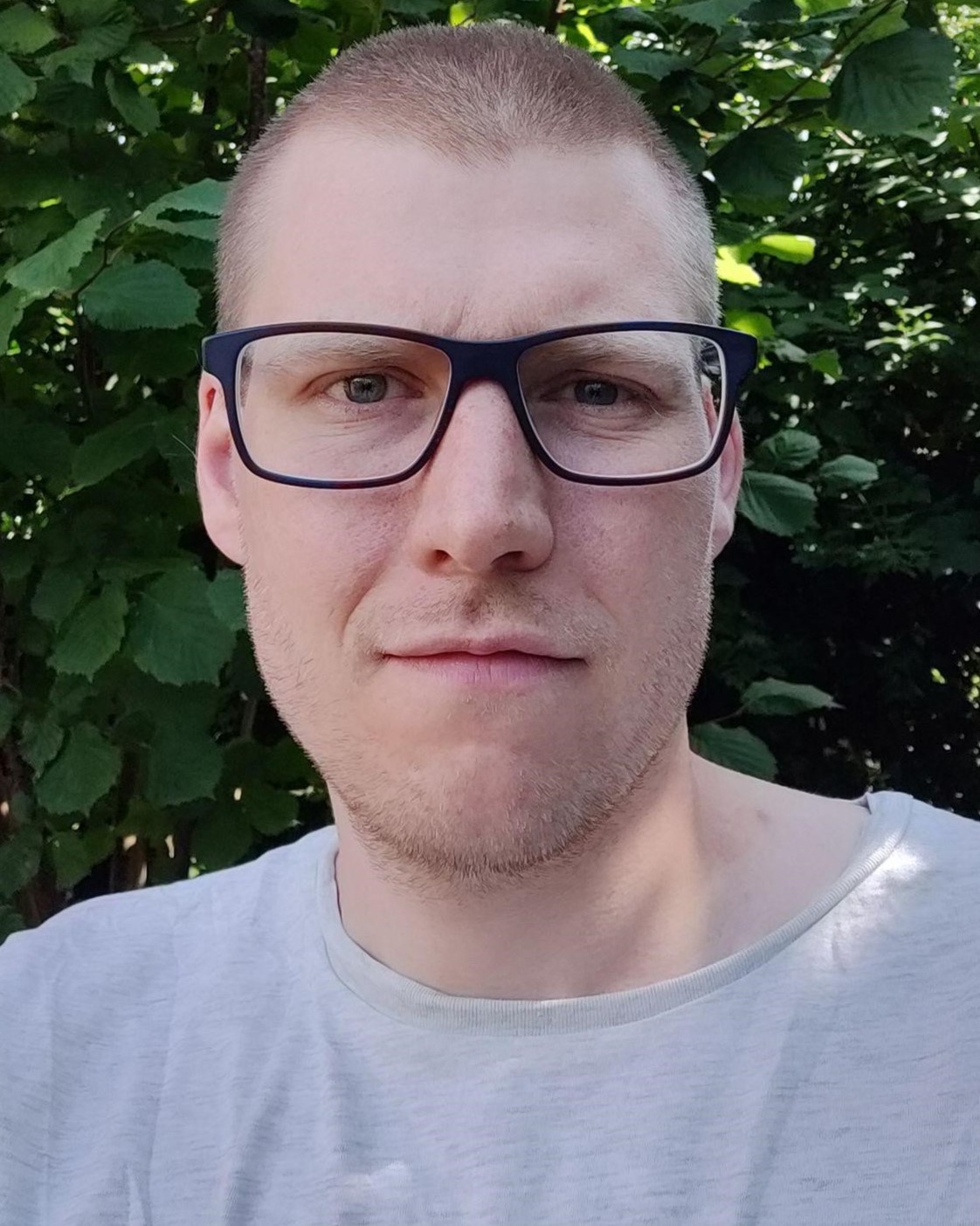}}]{Jon A. R. Liisberg}
received the B.Sc. degree in Mathematics from University of Copenhagen, Denmark in 2013. He received the M.Sc. and Ph.D. degrees in Mathematical modeling and computation from Technical University of Denmark in 2015 and 2019. The Ph.D. was jointly funded by Innovation Fund Denmark and Watts A/S. He is currently employed as a Data Scientist by Watts A/S, with focus on analysis and modeling of residential utility consumption. 
\end{IEEEbiography}

\vskip 0pt plus -1fil

\begin{IEEEbiography}[{\includegraphics[width=1in,height=1.25in,clip,keepaspectratio]{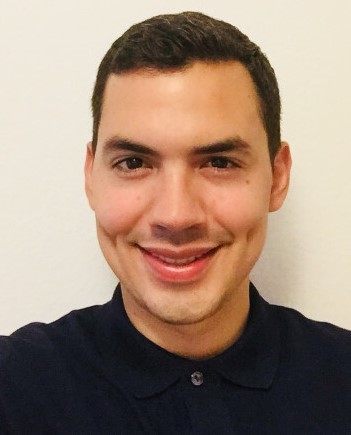}}]{Julián Lemos-Vinasco}
received the B.Sc. degree in industrial engineering from Universidad del Valle, Cali, Colombia in 2012, and the M.Sc. and Ph.D. degrees in mathematical modeling and computing from the Technical University of Denmark, Kongens Lyngby, Denmark, in 2017 and 2022, respectively. He is currently employed as a Data Scientist by Watts A/S as part of the company’s R\&D section. His main research areas include forecasting and optimisation tools for home energy management systems (HEMS), as well as machine learning application to residential utility consumption.
\end{IEEEbiography}







\end{document}